# CONSTRUCTING *OPERA SERIA* IN THE IBERIAN COURTS: METASTASIAN REPERTOIRE FOR SPAIN AND PORTUGAL[1]

# *CONSTRUYENDO* OPERA SERIA *EN LAS CORTES DE LA PENÍNSULA IBÉRICA: REPERTORIO DE METASTASIO PARA ESPAÑA Y PORTUGAL*

**Ana Llorens**
Instituto Complutense de Ciencias Musicales
allorens@iccmu.es
ORCID ID: 0000-0001-7290-9617

**Álvaro Torrente**
Universidad Complutense de Madrid
atorrente@iccmu.es
ORCID ID: 0000-0002-5830-183X.

**Abstract**

The exceptional reception of Pietro Metastasio's works during the eighteenth century, all over Europe and in the Iberian Peninsula in particular, is well documented. Due to that unparalleled success, it is possible to ascertain Spain and Portugal's participation in international, contemporary tastes and artistic webs, applicable to both composers and performers. However, this internationalisation needs to be nuanced, as some characteristics of the repertoire specifically written for the Peninsula indicate that their court audiences may have had expectations, both social and strictly musical, different from those of the public in opera theatres elsewhere in the continent. In this light, this article investigates in what ways the style of five composers in the international scene —Perez, Galuppi, Jommelli, Conforto, and Corselli— varied when commissioned to write *opera seria* for the Iberian courts.

The statistical analysis of fifteen settings especially written for the court theatres in Madrid and Lisbon, in comparison to the average data extracted from a corpus of 2,404 arias from 126 versions of a select number of Metastasian librettos, allows us to evaluate some particular usages regarding key, metre, tempo, and treatment of the vocal part. In this manner, through quantitative analysis, this article places eighteenth-century Iberian music production and consumption in the context of European *opera seria*, at the same time that it ultimately sets forth that its unique musical characteristics were also partly dependent on local musical customs, gender stereotypes, and personal idiosyncrasies alike.

**Resumen**

La excepcional acogida de las obras de Pietro Metastasio en el siglo XVIII, en Europa y en la península ibérica en particular, está ampliamente documentada. Debido a este éxito sin parangón, es posible afirmar que España y Portugal participaron de los gustos y redes artísticas internacionales del momento. Sin embargo, esta internacionalización ha de ser matizada, ya que el repertorio escrito específicamente para la península muestra que las audiencias cortesanas peninsulares pudieron tener expectativas diferentes a las del público de otros teatros en el resto del continente. Por ello, este trabajo investiga en qué formas varió el estilo de cinco compositores de talla internacional —Perez, Galuppi, Jommelli, Conforto y Corselli— a la hora de enfrentarse a los encargos de *opera seria* para las cortes ibéricas.

El análisis estadístico de quince versiones compuestas específicamente para Madrid y Lisboa nos permite evaluar, en comparación con las tendencias generales de un corpus de 2.404 arias tomadas de 126 versiones de libretos metastasianos, algunos usos locales en cuanto a tonalidad, compás, tempo y tratamiento de la parte vocal. A través del análisis cuantitativo, este artículo sitúa a la producción y el consumo de música en el siglo XVIII peninsular dentro del contexto de la *opera seria* europea, al mismo tiempo que, en última instancia, propone que sus características musicales también dependieron en parte de las costumbres musicales y los estereotipos de género del público de las cortes peninsulares, así como de algunos rasgos específicos de los compositores y los cantantes.

---

[1] This investigation is a result of the Didone Project, which has received funding from the European Research Council (ERC) under the European Union's Horizon 2020 research and innovation programme, Grant agreement No. 788986. Ana Llorens' work has been carried out under a "Juan de la Cierva-Formación" contract, funded by Spain's Agencia Estatal de Investigación, Ministerio de Ciencia e Innovación, ref. FJC2018-035534-I. We want to thank José Máximo Leza and José María Domínguez for their careful revision of our article, to Eduardo García-Portugués for his guidance in the statistical analysis, as well as to David Cranmer and Cristina Fernandes for their advice on Portuguese bibliography. Further thanks are due to the two anonymous reviewers of the article, who made important suggestions and prevented several errors.

**Key words**

Iberian Peninsula, *opera seria*, Metastasio, Spain, Portugal, style, international, statistical analysis, corpus studies, gender.

**Palabras clave**

Península ibérica, *opera seria*, Metastasio, España, Portugal, estilo, internacional, análisis estadístico, *corpus studies*, género.

In the eighteenth century, the *drammi per musica* by Pietro Metastasio (1698–1782) became a fever that took Europe by storm during almost a hundred years. As the chief magistrate of Madrid would explain around 1785:

> There was hardly a young man […] that did not know and did not sing by heart the "Misero pargoletto", the "Padre perdona", the "Son regina", "Se tutti mali miei", etc.[2] This fancy spread, already transformed into fashion in the theatre boxes, in all private or home performances; the directors of music, the composers and the orchestras were granted […] constant applause and encouragement […]. The number of amateurs increased and thus one could abundantly find and people would [uninterruptedly] sing the Italian arias, recitatives, rondos, and cavatinas by the best composers from that country.[3]

Since his *Didone abbandonata*, first set to music by Domenico Natale Sarro in 1724, and well into the following century,[4] Metastasio's 26 librettos[5] were musicalised by an average of 50 different composers,[6] with cases, such as his *Artaserse*, with almost a hundred complete versions,[7] some of which having been preserved only partially.[8] Although some of his dramas were set to music in the nineteenth century, such as in Saverio Mercadante's four settings —*Didone* (1823), *Ipermestra* (1825), *Ezio* (1827), and *Adriano* (1828)—, the French Revolution seems to have marked a major turning point, potentially underlining a link between Metastasian *opera seria* and the Ancient Régime. In any case, versions from ca. 1785 onwards epitomise a departure from the conventions of eighteenth-century *opera seria* (shorter libretti, preference for two acts, no tripartite but slow-fast arias, fewer arias and more ensembles, etc.), and also feature fewer poetic texts by Metastasio himself.[9]

---

[2] The first two arias belong to Metastasio's *Demofoonte*, and the latter two to his *Didone abbandonata*.

[3] "Apenas había un joven […] que no supiera y cantase de memoria el "Misero pargoletto", el "Padre perdona", el "Son regina", "Se tutti mali miei", etc. Corrió, pues, este gusto, ya hecho gusto de moda por los estrados en todas las funciones particulares o caseras; los maestros de música, los compositores y las orquestas lograron […] continuados aplausos y fomento […] y así se encontraban y cantaban a porfía las arias italianas, los recitados, los rondós y las cavatinas de los mejores compositores de aquel país". As quoted in Juan José Carreras and José Máximo Leza, "La recepción española de Metastasio durante el reinado de Felipe V (ca. 1730–1746)", in *Pietro Metastasio – uomo universale (1698–1782)*, ed. Andrea Sommer-Mathis and Elisabeth Theresia Hilscher (Viena: Verlag der Österreichischen Akademie der Wissenschaft, 2000), p. 254. The original source is José Antonio Armona y Murga, *Memorias cronológicas sobre el teatro en España (año de 1785)*, ed. Emilio Palacios Fernández, Joaquín Álvarez Barrientos and María del Carmen Sánchez García (Vitoria: Diputación Foral de Álava, 1988), p. 273. Unless otherwise stated, all translations are ours.

[4] For a comprehensive, yet not complete, catalogue of operas 1600–1900, see <http://corago.unibo.it/>.

[5] For a critical edition of Metatastasio's *drammi per musica*, see the "Progetto Metastasio", directed by Anna Laura Bellina and Luigi Tessarolo (<http://www.progettometastasio.it/public/>).

[6] The most comprehensive, yet incomplete, list of musical settings of Metastasio's dramas is found in Don Neville, "Metastasio [Trapassi], Pietro", *Grove Music Online*, <https://www.oxfordmusiconline.com/> [consulted: 4/4/2021]. We also resort to further sources, quoted appropriately. Niccolò Jommelli is a case in point. He musicalised Metastasio's *Demofoonte* on four occasions (Padua, 1743; Milan, 1753; Stuttgart, 1764; and Naples, 1770), always composing music completely different from that of any of his previous settings. For an analysis of the versions, see Tarcisio Balbo, "I quattro *Demofoonte* di Niccolò Jommelli: uno sguardo d'assieme", in *Demofoonte come soggetto per il dramma per musica: Johann Adolph Hasse ed altri compositori del Settecento*, ed. Milada Jonášová and Tomislav Volek (Prague: Academia, 2020), pp. 167-188. The same happens with *Artaserse*, for which he wrote three scores (Rome, 1749; Mannheim, 1751; and Stuttgart, 1756). However, the notions of authorship and "new version" in these settings are still unclear: while the *New Grove* lists two, Corago counts three authorised versions. See Marita P. McClymonds, Paul Cauthen, Wolfgang Hochstein, and Maurizio Dottori, "Jommelli [Jomelli]", *Grove Music Online,* <https://www.oxfordmusiconline.com/> [consulted: 24/02/2021].

[7] We have identified 98 different versions of *Artaserse*, 20 more than Neville for *New Grove*; these will eventually be published in a detailed catalogue.

[8] Some composers wrote separate arias as well, normally on the most famous verses of the dramas. An example is "Misero pargoletto" from *Demofoonte*, set to music by Mozart (KV 73e) and Schubert (D. 42). For a complete catalogue of *Demofoonte* settings, see Ana Llorens, Gorka Rubiales and Nicola Usula, "Operatic sources for *Demofoonte*: Librettos and scores after Metastasio's 'figliuolo'", in *Demofoonte come soggetto per il dramma per musica: Johann Adolph Hasse ed

Given the scant presence of the operatic genre in Spain and Portugal still around the turn of the eighteenth century,[10] one could well expect that the Iberian influence of Metastasian drama was, if not weak, at least late. Yet the Iberian Peninsula as a whole was not a European exception and did not escape the fashion. As Stein and Leza explain:

> Opera productions in the Hispanic world formed part of what might be termed the international matrix […]. The travels of Italian libretti […] and their revival or adaptation beyond Italy, broadened the reach of opera and the pan-European aesthetic, even as the travels of Spanish diplomats and Italian impresarios, as well as Italian and Spanish singers and composers, traced an active network through and across three continents.[11]

And the same could be said of Portugal and its territories, where Brito writes of an Italian "invasion" from 1719 and particularly in the 1730s.[12]

The first Metastasian production in the Iberian Peninsula was the performance in 1734 of an *Artaserse* by an unknown composer, "ejecutada en el italiano" in the palace of the Viceroy of Valencia to celebrate Elisabeth Farnese's birthday, only four years after the first version of this libretto.[13] Vernacular adaptations, transforming operas into *comedias*, started to be performed in Madrid from 1736,[14] the same year of the first performances of Metastasio's *drammi per musica* in Lisbon.[15] The poet's fortune continued with several operatic performances in Madrid from 1738, including the first court production, Corselli's *Alessandro nell'Indie*.[16] Apropos

---

*altri compositori del Settecento*, ed. Milada Jonášová and Tomislav Volek (Prague: Academia, 2020), pp. 271-317.

[9] An example is Saverio Mercadante's *Didone abbandonata*. In it, only four —out of the fourteen closed numbers in the score— Metastasian aria texts are set to music: "Son regina e sono amante" (Act 1-Scene 4), "Quando saprai chi sono" (A1-S5; originally an aria, here a duet), "Ogni amator supone" (A2-S8), and "Ah non lasciarmi, no" (A2-S9; originally an aria, here a duet).

[10] See Louise K. Stein and José Máximo Leza, "Opera, genre and context in Spain and its American colonies", in *The Cambridge Companion to Eighteenth-Century Opera*, ed. Anthony R. DelDonna and Pierpaolo Polzonetti (New York: Cambridge University Press, 2009), p. 244.

[11] Stein and Leza, "Opera, genre and context in Spain", p. 268. See also Manuel Carlos de Brito, "Ópera e teatro musical em Portugal no século XVIII", in *Teatro y música en España (siglo XVIII). Actas del Simposio Internacional Salamanca 1994*, ed. Rainer Kleinertz (Kassel and Berlin: Reichenberger, 1996), p. 178.

[12] Brito, "Ópera e teatro musical em Portugal", p. 179.

[13] See Emilio Cotarelo y Mori, *Orígenes y establecimiento de la ópera en España hasta 1800* (Madrid: Tip. de la "Revista de Archivos, Bibliotecas y Museos", 1917), p. 270. For a discussion of Metastasio's Spanish adaptations during the first half of the century, see Carreras and Leza, "La recepción española de Metastasio", esp. pp. 255-262; and Leza, "Metastasio on the Spanish Stage". For an investigation of such adaptations in the second half of the eighteenth century, see Patrizia Garelli, "Metastasio y el melodrama italiano", in *El teatro europeo en la España del siglo XVIII*, ed. Francisco Lafarga (Lleida: Edicions Universitat de Lleida, 1997), pp. 127-138. For a study of commercial Italian opera in the 1730s and the 1740s in Spain, see José Máximo Leza, "Francesco Corradini y la introducción de la ópera en los teatros comerciales de Madrid (1731–1749)", *Artigrama*, 12 (1996–1997), pp. 123-146. For a discussion of the phenomenon in Portugal in the 1730s, see Brito, "Ópera e teatro musical em Portugal", p. 180.

[14] Maria Grazia Profeti, "El espacio del teatro y el espacio del texto: Metastasio en España en la primera mitad del siglo XVIII", in *La ópera en España e Hispanoamérica*, ed. Emilio Casares and Álvaro Torrente (Madrid: Instituto Complutense de Ciencias Musicales, 2001), p. 290; Reiner Kleinertz, *Grundzüge des spanischen Musiktheaters im 18. Jahrhundert* (Kassel: Reichenberger, 2003), pp. 77-112.

[15] Portuguese adaptations of Metastasio's *drammi per musica* started in the 1740s and remained popular in public theatres until the end of the century. See David Cranmer, *Peças de um mosaico: temas da história da música referentes a Portugal e ao Brasil* (Lisboa: Edições Colibri, 2017), particularly the chapters "A música no teatro popular português setecentista: à procura de um paradigma" and "A ópera e a comédia em língua portuguesa durante o reinado de D. José I".

[16] With the court premiere of Corselli's *Alessandro nell'Indie* in the Buen Retiro theatre. Apparently, it had already been performed in the marquis of Santiago's house, yet in translation. See Carreras and Leza, "La recepción española de Metastasio", p. 254.

three-act *drammi per musica*, court productions would comprise the 42% of the Metastasian consumption in the Peninsula.

The use of the renovated Buen Retiro theatre[17] and, especially, the arrival in Madrid of the first professional companies of Italian singers[18] marked the start of a couple of flourishing years during which Italian virtuosos toured in Spain too, singing a repertoire similar to the one heard in the rest of Europe. Both in the courts and elsewhere in the two countries, there was a suspension in operatic productions: in Spain it coincided with the end of the reign of Felipe V (1738–1746), with Corselli's *Achille in Sciro* (1744), premiered on the occasion of a royal wedding, as the only production between 1739 and the first opera under Fernando VI in 1747;[19] in Portugal the interruption was caused by João V's prohibition of all theatrical spectacles in 1742, in force until his death in 1750.[20]

The enthronements of Fernando VI (1746–1759) in Spain and of José I (1750–1777) in Portugal resulted in a climax for the operatic scene.[21] The Bourbon king appointed Farinelli, a close friend of Metastasio's, as artistic director of the court spectacles. As Table 1a shows, Spanish court opera resumed activities in 1747 with *La clemenza di Tito* (music by Corselli, Corradini, and Mele), commissioning from 1751 at least one, and sometimes up to three, new versions of a Metastasian libretto every year during a decade. In fact, some of the works in this "Metastasian cycle" were abridged by the poet himself, who "knew how to adapt his texts to the local circumstances and needs".[22] In Portugal (Table 1b) a similarly frenetic operatic life in the court had to wait until 1752, yet it remained steady with an average of two yearly new productions and several revivals until 1755. In fact, the Braganza sovereign's enthusiasm is undeniable: short after the decease of his father João V he was "already involved in negotiations to hire some of the best Italian opera singers then available […] sparing no expense to secure" the services of the best ones, including Gizziello and Raaf.[23] The Portuguese court hired David Perez as director of music in March 1751,[24]

---

[17] The Coliseo del Buen Retiro underwent two renovations in this period, one in 1738 and a major rebuilding in 1747, with a view to adapting it to the performance of Italian opera. See Margarita Torrione, "El Coliseo del Buen Retiro: Memoria de una arquitectura desaparecida", in *España festejante. El siglo XVIII*, ed. Margarita Torrione (Málaga: Centro de Ediciones de la Diputación de Málaga, 2000), pp. 295-322.

[18] Juan José Carreras, "Terminare a schiaffoni: La primera compañía de ópera italiana en Madrid (1738/9)", *Artigrama*, 12 (1996–1997), pp. 99-121.

[19] Carreras situates the performances of Corselli's *Alessandro nell'Indie* (1738), *Farnace* (1739), and *Achille in Sciro* (1744) at the same level of importance, although with different functions, for the Spanish court. See Juan José Carreras, "En torno a la introducción de la ópera de corte en España: *Alessandro nell'Indie* (1738)", in *España festejante. El siglo XVIII*, ed. Margarita Torrione (Málaga: Centro de Ediciones de la Diputación de Málaga, 2000), p. 325; and Reinhard Strohm, "Francesco Corselli's operas for Madrid", in *Teatro y música en España (siglo XVIII). Actas del Simposio Internacional Salamanca 1994*, ed. Rainer Kleinertz (Kassel and Berlin: Reichenberger, 1996), pp. 79-106.

[20] See Manuel Carlos de Brito, *Opera in Portugal in the Eighteenth Century* (Cambridge: Cambridge University Press, 1989), p. 22.

[21] For a comprehensive chronology of operas performed in Portugal in the eighteenth century, see Brito, *Opera in Portugal*, pp. 121-175. Thanks to later research this list can be completed with further titles.

[22] Andrea Sommer-Mathis, "Entre Viena y Madrid, el tandem Metastasio-Farinelli: dirección escénica y dirección artística", in *España festejante. El siglo XVIII*, ed. Margarita Torrione (Málaga: Centro de Ediciones de la Diputación de Málaga, 2000), p. 384. Metastasio reduced the text of four librettos for the Spanish court, including *Alessandro nell'Indie*, originally written for Rome in 1730. See Carlos Broschi "Farinelli", in *Fiestas reales*, ed. Antonio Bonet Correa and Antonio Gallego (Madrid: Patrimonio Nacional, 1991), p. 185. The other reduced libretti were *Adriano in Siria*, *Didone abbandonata*, and *Semiramide*. On the latter, See Javier Gutiérrez Carou, "The textual evolution of Metastasio's *Semiramide*: Aesthetic transformation and proportional identity", in *Mapping Artistic Networks: Eighteenth-Century Italian Theatre and Opera across Europe*, ed. Tatiana Korneeva (Turnhout, Brepols, 2021); and Javier Gutiérrez Carou, "Evoluzione musicale e struttura letteraria del melodramma metastasiano nel contesto settecentesco: pezzi chiusi e protagonismo plurimo", *Rivista di letteratura italiana*, 38/3 (2020), pp. 63-90. See also Garelli, "Metastasio y el melodrama italiano", p. 129; José Máximo Leza, "'Al dulce estilo de la culta Italia': ópera italiana y zarzuela española", in *Historia de la música en España e Hispanoamérica*, vol. 4*: La música en el siglo XVIII*, ed. José Máximo Leza (Madrid: Fondo de Cultura Económica, 2004), p. 318; and Sommer-Mathis, "Entre Viena y Madrid", pp. 387-390.

[23] Brito, *Opera in Portugal*, p. 24.

[24] Since his appointment in 1752, Perez set to music seven Metastasian *drammi* for the Portuguese court: *Demofoonte* (1752), *L'eroe cinese* and *Olimpiade* (1753), *Adriano in Si-*

unsuccessfully competed with other European courts to attract Niccolò Jommelli in 1753,[25] and built a new theatre —inaugurated in March 1755 with Mazzoni's *Clemenza*— that costed "him two million a year".[26] All of this coincided with the general peak of European opera on Metastasio's texts in the 1750s. Not surprisingly, Metastasio wrote his only *opera seria* libretto specifically for premiere in the Peninsula (*Nitteti*, 1756) in these years, which further attests to the Spanish court's predilection for his poetry.[27]

In the second half of the 1750s, two events marked a dramatic halt of opera productions in the Peninsula while in the rest of Europe the trend was still burgeoning: the Lisbon earthquake on 1 November 1755 and the decease of Queen Bárbara de Braganza three years later. As the following monarchs had different predilections, court opera productions became less frequent under Carlos III[28] and IV of Spain (1759-1788 and 1788-1808, respectively) and José I of Portugal (1750-1777), coinciden-tally with the new attraction for the *opera buffa*[29] and the courtly promotion of French dramas.[30] In Lisbon, in fact, no Metastasian opera was heard in eight years, until the performance of a *Temistocle* by an unknown composer in 1763. As Brito puts it, "the earthquake brought operatic activity to a complete standstill. It was only in 1763 that it was again resumed on a somewhat more modest scale, at least as far as singers were concerned",[31] most of them serving not only the opera house but also the Royal Chapel, pointing to a more economical approach. However, the king insisted on commissioning Jommelli new *seria* works from 1769.[32]

Contrary "to what happened in Spain, court opera would remain operational", although scarce, in Portugal "until almost the end of the century",[33] even after the prohibition of female singers since 1775 and during Maria I's (1777-1792) reign. Conversely, court opera was eradicated from the Spanish court by Charles III, and no performance of a Metastasian drama took place again in Madrid until the 1790, when several revivals by fashionable Italian composers were staged in public theatres.[34] This followed the trend in the previous decades (1760-1785) to perform revivals and *pasticcios* in commercial theatres in Barcelona and Cádiz, and, less frequently, in Palma, Jerez, and Valencia. In Spain court and commercial theatres interacted throughout the century, with techniques from the Spanish dramaturgy being adopted when translating, shortening, or adapting Metastasio's librettos, and with the *zarzuela* and the *comedia* being influenced by the *drama per musica* and the *drama giocoso*.[35] Nothing of this happened in Portugal,

---

*ria* and *Ipermestra* (1754), *Alessandro nell'Indie* (1755), and *Demetrio* (1765). Also, his *Didone abbandonata* from 1751 was revived in 1753 and his 1748 *Artaserse* in 1754. Long after the earthquake, from 1765, five of his operas were revived, including *Didone abbandonata*, *Semiramide*, and *Zenobia* (1765), *Demetrio* (1766, 1768), *Demofoonte* (1772), and *Alessandro* (1808), as he devoted his efforts mainly to church music.

[25] Saverio Mattei, Jommelli's close friend and first biographer, explains that three European courts were fighting in 1753 to get the services of Jommelli. At last, "nella gara vinse quella da Stugard; e Jommelli risolse di preferirla per la delicatezza del gusto del Duca di Wittemberg". See Saverio Mattei, "Elogio del Jommelli osia il progeresso della poesia, e della musica teatrale", in *Opere del Signor Abate Pietro Metastasio Romano* (Napoli: Fratelli de Bonis, 1784), vol. XIII, pp. liii-cxx; quote on p. lxxv.

[26] Chevalier des Courtils' testimony, as cited in Brito, *Opera in Portugal*, p. 27.

[27] On this occasion, the music was written by Nicola Conforto (1718–1793), opera composer to the Spanish court since that year. See Gian Giacomo Stiffoni, "Per una biografia del compositore napoletano Nicola Conforto (Napoli, 1718–Madrid, 1793). Documenti d'archivio, libretti conservati nella Biblioteca Nacional di Madrid, fonti musicali manoscritte e a stampa", *Fonti Musicali Italiane*, 4 (1999), pp. 7-54.

[28] Gian Giacomo Stiffoni, "La ópera de corte en tiempos de Carlos III (1759-1788)", in *La ópera en España e Hispanoamérica*, ed. E. Casares y A. Torrente (Madrid: Instituto Complutense de Ciencias Musicales, 2002), pp. 317-341.

[29] Brito, *Opera in Portugal*, p. 51; Stein and Leza, "Opera, genre and context in Spain", pp. 261-262; and Stiffoni, "La ópera de corte en tiempos de Carlos III".

[30] José Máximo Leza, "El mestizaje ilustrado: influencias francesas e italianas en el teatro musical madrileño (1760–1780)", *Revista de Musicología*, 32/2 (2009), pp. 503-546.

[31] Brito, *Opera in Portugal*, p. 31.

[32] Brito, *Opera in Portugal*, p. 40.

[33] Brito, "Ópera e teatro musical em Portugal", p. 183.

[34] Brito, "Ópera e teatro musical em Portugal", p. 185.

[35] Leza, "El mestizaje ilustrado", p. 506. See also José Máximo Leza, "L'aria col da capo nella zarzuela spagnola a metà del settecento", *Musica e Storia*, 16/3 (2008), pp. 587-613; Leza, "'Al dulce estilo de la culta Italia'", pp. 320ff; and Antonio Soriano Santacruz, "Aquiles en Sciro de Ramón de la Cruz y Blas de Laserna. La pervivencia de Metastasio en las comedias populares en el Madrid de la Ilustración", *Scherzo*, 360 (2020), pp. 82-85.

where bourgeoisie and royalty remained detached from one another. The removal of Maria I from the throne in 1792 marked the end of court opera in Portugal, coincidently with the construction of the new Teatro São Carlos in 1793, funded by the rising bourgeoisie.[36]

In line with the general trend in Europe, after the 1790s a number of Metastasio's operas were also revived in Portugal.[37] Interestingly, three new versions were premiered in Lisbon: Marino's *Didone abbandonata* (1798) and Marcos Portugal's *Artaserse* (1806) and *Demofoonte* (1808).[38] Productions decreased until the 1820s, when Metastasio's librettos were heard in the Peninsula for the last time: Cimarosa's *Clemenza* in Lisbon (1821) and Mercadante's *Didone* in Barcelona (1826). Tables 1a and 1b summarise the 181 documented productions of Metastasio's *drammi* in Portugal (61) and Spain (120).[39]

| Year | Date | Opera | Composer(s) | City | Type | Orig. version | Ref. |
|---|---|---|---|---|---|---|---|
| 1734 |  | *Araserse* | Unknown | Valencia | revival? |  | Cotarelo |
| 1738 | 9 Feb | *Demetrio* | Hasse, other | Madrid | revival | 1732 | Leza |
| 1738 | 21 May | *Demofoonte* | Schiassi, other | Madrid | pasticcio? |  | Leza |
| 1738 | 25 Oct | *Artaserse* | Hasse, Vinci | Madrid | pasticcio? |  | Leza |
| 1738 | 9 May | *Alessandro* | Corselli | Madrid-CT | premiere |  | Leza |
| 1738 | 8 Jul | *Alessandro* | Corselli | Madrid-CT | revival | 1738 | Leza |
| 1738 | 19 Dec | *Alessandro* | Corselli | Madrid-CT | revival | 1738 | Leza |
| 1739 | 29 Mar | *Siroe* | Hasse | Madrid | revival | 1733 | Leza |
| 1739 | 14 May | *Clemenza* | Hasse | Madrid | revival | 1735 | Leza |
| 1744 | 8 Dec | *Achille* | Corselli | Madrid-CT | premiere |  | Leza |
| 1747 | 20 Jan | *Clemenza* | Corselli, Corradini, Mele | Madrid-CT | premiere |  | Leza |
| 1749 | 6 Jan | *Artaserse* | Mele, other | Madrid-CT | pasticcio? |  | Leza |
| 1749 | 6 Apr | *Artaserse* | Mele, other | Madrid-CT | pasticcio? |  | Casanova |
| 1749 | 18 Dec | *Demofoonte* | Galuppi, Mele | Madrid-CT | premiere |  | Leza |
| 1749 | 31 Dec | *Artaserse* | Mele, other | Madrid-CT | pasticcio? |  | Casanova |
| 1750 | 4 Dec | *Alessandro* | Scolari | Barcelona | revival | 1750 | Alier |
| 1750 | 1 Jan | *Artaserse* | Mele, other | Madrid-CT | pasticcio? | 1749? | Leza |
| 1750 | 21 Feb | *Demofoonte* | Galuppi, Mele | Madrid-CT | revival | 1749 | Leza |
| 1751 | 30 May | *Demofoonte* | Galuppi | Barcelona | revival | 1749 | Alier |
| 1751 | 23 Sep | *Siroe* | Various | Barcelona | pasticcio |  | Alier |

---

[36] Brito, "Ópera e teatro musical em Portugal", pp. 185-186.
[37] In Spain, the model of Italian opera resumed in the Caños del Peral theatre in 1786 thanks in part to a strong and necessary publicity and divulgation campaign. Leza, "El mestizaje ilustrado", p. 523.
[38] Interestingly, Portugal's new setting of *Demofoonte* was commissioned by French military authorities to celebrate the birthday of emperor Napoleon on 15 August, as explained in Cranmer, *Peças de un mosaico*, p. 199. Lucio Tufano reveals Napoleon's inclination for this opera, particularly for its most famous aria "Misero pargoletto". See Lucio Tufano, "Sulla fortuna di *Misero pargoletto*: materiali e ipotesi", in *Demofoonte come soggetto per il dramma per musica: Johann Adolph Hasse ed altri compositori del Settecento*, ed. Milada Jonášová and Tomislav Volek (Prague: Academia, 2020), pp. 87-108.
[39] Tables 1a and 1b include *drammi per musica* only and not Iberian adaptations of Metastasio's Italian dramas. For an account of these, see especially José Máximo Leza, "Metastasio on the Spanish stage: operatic adaptations in the public theatres of Madrid in the 1730s", *Early Music*, 26/4 (1998), pp. 623-631.

| Year | Date | Opera | Composer(s) | City | Type | Orig. version | Ref. |
|---|---|---|---|---|---|---|---|
| 1751 | 14 Feb | *Demofoonte* | Galuppi, Mele | Madrid-CT | revival | 1749 | Leza |
| 1751 | 23 Sep | *Demetrio* | Jommelli | Madrid-CT | premiere | | Leza |
| 1752 | 30 May | *Didone* | Scolari | Barcelona | premiere | | Alier |
| 1752 | | *Alessandro* | Scolari | Barcelona | revival | 1750 | Alier |
| 1752 | 1 Jan | *Demetrio* | Jommelli | Madrid-CT | revival | 1751 | Leza |
| 1752 | 20 Jan | *Demofoonte* | Galuppi, Mele | Madrid-CT | revival | 1749 | Leza |
| 1752 | 1 Feb | *Demetrio* | Jommelli | Madrid-CT | revival | 1751 | Leza |
| 1752 | 11 Apr | *Demofoonte* | Galuppi, Mele | Madrid-CT | revival | 1749 | Leza |
| 1752 | 23 Sep | *Didone* | Galuppi | Madrid-CT | premiere | | Leza |
| 1752 | 19 Dec | *Didone* | Galuppi | Madrid-CT | revival | 1752 | Leza |
| 1752 | 25 Dec | *Siroe* | Conforto | Madrid-CT | premiere | | Leza |
| 1753 | Oct | *Didone* | Scolari | Barcelona | revival | 1752 | Alier |
| 1753 | | *Il re pastore* | Unknown | Barcelona | ? | | Alier |
| 1753 | 2 Feb | *Demetrio* | Jommelli | Madrid-CT | revival | 1751 | Leza |
| 1753 | 4 Feb | *Siroe* | Conforto | Madrid-CT | revival | 1752 | Casanova |
| 1753 | 6 Jun | *Siroe* | Conforto | Madrid-CT | revival | 1752 | Leza |
| 1753 | 23 Sep | *Semiramide* | Jommelli | Madrid-CT | premiere | | Leza |
| 1753 | 4 Dec | *Semiramide* | Jommelli | Madrid-CT | revival | 1753 | Casanova |
| 1753 | 18 Dec | *Didone* | Galuppi | Madrid-CT | revival | 1752 | Leza |
| 1753 | 4 Dec | *Alessandro* | Scolari | Puerto S. María | revival | 1750 | Corago |
| 1754 | Spring | *Ezio* | Unknown | Barcelona | ? | | Alier |
| 1754 | 19 May | *Semiramide* | Unknown | Barcelona | ? | | Corago |
| 1754 | Oct | *Artaserse* | Ferrandini | Barcelona | revival | 1739 | Alier |
| 1754 | 20 Jan | *Siroe* | Conforto | Madrid-CT | revival | 1752 | Casanova |
| 1754 | 14 Apr | *Didone* | Galuppi | Madrid-CT | revival | 1752 | Casanova |
| 1754 | 23 Sep | *L'eroe cinese* | Conforto | Madrid-CT | premiere | | Leza |
| 1754 | 4 Dec | *L'eroe cinese* | Conforto | Madrid-CT | revival | 1754 | Leza |
| 1754 | 18 Dec | *Didone* | Galuppi | Madrid-CT | revival | 1752 | Leza |
| 1755 | | *Clemenza* | Various | Barcelona | pasticcio? | | Alier |
| 1755 | 30 May | *Achille* | Jommelli? | Barcelona | revival | 1749 | Corago |
| 1755 | 23 Sep | *L'eroe cinese* | Bonno | Barcelona | revival | 1752 | Corago |
| 1755 | 6 Jan | *L'eroe cinese* | Conforto | Madrid-CT | revival | 1754 | Leza |
| 1755 | 12 Jan | *Didone* | Galuppi | Madrid-CT | revival | 1752 | Leza |
| 1755 | 30 Mar | *Didone* | Galuppi | Madrid-CT | revival | 1752 | Casanova |
| 1755 | 1 Apr | *L'eroe cinese* | Conforto | Madrid-CT | revival | 1754 | Leza |
| 1755 | 6 Apr | *Didone* | Galuppi | Madrid-CT | revival | 1752 | Leza |
| 1755 | 23 Sep | *Demofoonte* | Galuppi, Mele? | Madrid-CT | revival | 1749? | Leza |
| 1755 | 18 Dec | *Demofoonte* | Galuppi, Mele? | Madrid-CT | revival | 1749? | Carreras |
| 1756 | 30 May | *Temistocle* | Various | Barcelona | pasticcio? | | Alier |
| 1756? | | *Semiramide* | Ferrandini | Madrid-CT? | premiere? | | Libby |
| 1756 | | *Il re pastore* | Ferrandini | Madrid-CT? | premiere? | | Libby |
| 1756 | 23 Sep | *Nitteti* | Conforto | Madrid-CT | premiere | | Leza |
| 1756 | 4 Dec | *Nitteti* | Conforto | Madrid-CT | revival | 1756 | Leza |

| Year | Date | Opera | Composer(s) | City | Type | Orig. version | Ref. |
|---|---|---|---|---|---|---|---|
| 1757 | 2 Jan | *Il re pastore* | Mazzoni | Madrid-CT | premiere | | Leza |
| 1757 | 10 Apr | *Nitteti* | Conforto | Madrid-CT | revival | 1756 | Leza |
| 1757 | 23 Sep | *Adriano in Siria* | Conforto | Madrid-CT | premiere | | Leza |
| 1757 | 4 Dec | *Nitteti* | Conforto | Madrid-CT | revival | 1756 | Leza |
| 1757 | 25 Dec | *Adriano* | Conforto | Madrid-CT | premiere | | Leza |
| 1758 | 28 Mar | *Adriano* | Conforto | Madrid-CT | revival | 1757 | Casanova |
| 1758 | 31 Mar | *Nitteti* | Conforto | Madrid-CT | revival | 1756 | Casanova |
| 1758 | 31 May | *Nitteti* | Conforto | Madrid-CT | revival | 1756 | Leza |
| 1760 | 10 Jul | *Antigono* | Durán | Barcelona | premiere | | Alier |
| 1761 | | *Siroe* | Various | Cádiz | pasticcio? | | Corago |
| 1762 | Sep | *Alessandro* | Scolari | Barcelona | revival | 1750 | Alier |
| 1762 | 4 Nov | *Temistocle* | Durán | Barcelona | premiere | | Alier |
| 1762 | | *Zenobia* | Perez | Cádiz | revival | 1751 | Corago |
| 1762 | 24 Dec | *Olimpiade* | Galuppi | Cádiz | revival | 1747 | Corago |
| 1763 | | *Didone* | Scolari | Barcelona | revival | 1753 | Alier |
| 1763 | 15 Jan | *Artaserse* | Piccinni | Barcelona | revival | 1751 | Corago |
| 1763 | 6 May | *Adriano* | Sciroli | Barcelona | premiere | | Alier |
| 1763 | 22 Aug | *Demofoonte* | Galuppi | Barcelona | revival | 1758? | Alier |
| 1763 | 25 Sep | *Catone* | Various | Barcelona | pasticcio? | | Alier |
| 1764 | May | *Demofoonte* | Galuppi | Cádiz | revival | 1749 | Kleinertz |
| 1764 | Sep | *Antigono* | Various | Cádiz | pasticcio? | | Corago |
| 1764 | Dec | *Alessandro* | Perez | Cádiz | revival | 1755 | I-Bc |
| 1765 | | *Siface* | Various | Barcelona | pasticcio? | | Corago |
| 1766 | 15 Aug | *Zenobia* | Unknown | Barcelona | ? | | Corago |
| 1766 | Autumn | *Attilio Regolo* | Jommelli | Barcelona | revival | 1753 | Alier |
| 1766 | 4 Nov | *Clemenza* | Valentini | Barcelona | revival | 1753 | Alier |
| 1767 | | *Didone* | Unknown | Barcelona | ? | | Corago |
| 1767 | 5 May | *Ezio* | Traetta | Barcelona | revival | 1754? | Alier |
| 1767 | 4 Nov | *Alessandro* | Scolari | Barcelona | revival? | 1750 | Alier |
| 1767 | | *Artaserse* | Unknown | Palma | ? | | Corago |
| 1767 | | *Demofoonte* | Galuppi | Palma | revival | 1749 | Corago |
| 1768 | 4 Nov | *Didone* | Unknown | Cádiz | ? | | Corago |
| 1769 | 4 Nov | *Olimpiade* | Piccinni | Barcelona | revival | 1768 | Alier |
| 1769 | 9 Dec | *Zenobia* | Various | Cádiz | pasticcio? | | Corago |
| 1769 | 20 Jan | *Artaserse* | Sacchini | Valencia | revival | 1768 | Corago |
| 1769 | 4 Nov | *Demofoonte* | Various | Valencia | pasticcio? | | Corago |
| 1770 | 3 Feb | *Nitteti* | Unknown | Cádiz | ? | | Díez |
| 1770 | Carniv. | *Achille* | Monza | Cádiz | revival | 1764 | Díez |
| 1770 | Carniv. | *Demofoonte* | Bernasconi | Cádiz | revival | 1756? | Díez |
| 1770 | 1 Jul | *Semiramide* | Galuppi | Jerez | revival | 1749 | Corago |
| 1770 | | *La clemenza* | Valentini? | Barcelona | revival? | 1753 | Alier |
| 1772 | | *Semiramide* | Unknown | Barcelona | ? | | Corago |
| 1772 | 20 Jan | *Demetrio* | Piccinni | Barcelona | revival | 1769 | Alier |

| Year | Date | Opera | Composer(s) | City | Type | Orig. version | Ref. |
|---|---|---|---|---|---|---|---|
| 1772 | 4 Nov | *Demofoonte* | Unkown | Barcelona | ? | | Alier |
| 1773 | 25 Aug | *Ciro riconosciuto* | Various | Cádiz | pasticcio | | Díez |
| 1774 | Carniv. | *Antigono* | Perez | Cádiz | premiere? | ? | Díez |
| 1774 | 14 Aug | *Semiramide* | Galuppi | Valencia | revival | 1749 | Corago |
| 1777 | 9 Dec | *Trionfo di Clelia* | Jommelli? | Barcelona | revival? | 1774 | Alier |
| 1778 | | *Olimpiade* | Anfossi | Cádiz | revival | 1774 | Díez |
| 1785 | | *L'eroe cinese* | Cimarosa | Barcelona | revival | 1782 | Corago |
| 1790 | | *Demofoonte* | Various | Madrid? | pasticcio? | | |
| 1791 | | *Didone* | Andreozzi | Madrid | revival | 1784 | Corago |
| 1792 | | *Alessandro* | Caruso | Madrid | revival | 1787 | Corago |
| 1792 | | *Didone* | Sarti | Madrid | revival | 1782 | Corago |
| 1793 | | *Ipermestra* | Paisiello, other | Madrid | revival | 1791 | Corago |
| 1794 | | *Demetrio* | Guglielmi, other | Madrid? | pasticcio? | | Kleinertz |
| 1826 | | *Didone* | Mercadante | Barcelona | revival | 1823 | Corago |

**Table 1a.** Performances of Metastasio's *drammi per musica* in Spain, 1734-1826. The main bibliographical reference or the library siglum for the manuscript source used is indicated in the last column.
CT = court theatre.[40]

---

[40] Two chronologically close performances of the same opera are considered independent productions when they take place more than 30 days apart, as there are many cases of performances in consecutive months separated by around one week. The references used are: Roger Alier y Aixalà, *L'òpera a Barcelona: orígens, desenvolupament i consolidació de l'òpera com a espectacle teatral a la Barcelona del segle XVIII* (Barcelona: Institut d'Estudis Catalans, 1990); Brito, *Opera in Portugal*; Corago: <http://corago.unibo.it/>; Juan José Carreras, "Farinelli's dream: Theatrical space, audience and political function of Italian court opera", in *Musiktheater im höfischen Raum des frühneuzeitlichen Europa*, ed. Margret Scharrer, Heiko Laß, and Matthias Müller (Heidelberg: Heidelberg University Publishing, 2020), Appendix II, p. 386; Teresa Casanova, "El intermezzo en la corte de España, 1738-1758", Ph.D. diss., Universidad Complutense de Madrid, 2019, Table 4.6, pp. 167-169; Cristina Díez Rodríguez, "Cádiz, centro operístico peninsular en la España de los siglos XVIII y XIX (1761-1830)", Ph.D. diss., Universidad Complutense de Madrid, 2015; Paul Joseph Jackson, "The operas of David Perez", Ph.D. diss., Stanford University, 1967; Kleinertz, *Grundzüge,* vol. 2; Leza, "'Al dulce estilo de la culta Italia'", Table 23, pp. 332-339; Dennis Libby, James L. Jackman and Rebecca Green, "Ferradini [Feradini, Ferrandini], Antonio", in *Grove Music Online*, <https://www.oxfordmusiconline.com/> [consulted: 17/03/2021]; António Jorge Marques and David Cranmer, "Portugal [Portogallo], Marcos [Marco] António (da Fonseca)", in *Grove Music Online*, <https://www.oxfordmusiconline.com/> [consulted: 27/03/2021]; Marita P. McClymonds, "Niccolò Jommelli: The last years", Ph.D. diss., University of California Berkeley, 1978. The only reference to the 1756 production of Ferrandini's *Semiramide* is in Libby; the manuscript source for the only preserved aria ("Caria ti lascio, addio", D-Dl Mus.3054-F-1,8) indicates "Teatro Reale di Madrid" and, since no libretto has been preserved either, we consider it doubtful.

| Year | Date | Opera | Composer(s) | City | Type | Orig. version | Ref. |
|---|---|---|---|---|---|---|---|
| 1736 | Jan–Feb | Alessandro | Schiassi | Lisbon | revival | 1734 | Brito |
| 1737 | May | Artaserse | Schiassi | Lisbon | premiere | | Brito |
| 1737 | May | Demofoonte | Schiassi | Lisbon | revival | 1734 | Brito |
| 1737 | May | Olimpiade | Unknown | Lisbon | ? | | Brito |
| 1737 | af. May | Siface | Leo | Lisbon | revival | 1737 | Corago |
| 1738 | Nov | Clemenza | Unknown | Lisbon | ? | | Brito |
| 1738 | Mar | Semiramide | Unknown | Lisbon | ? | | Brito |
| 1738 | Mar | Siroe | Unknown | Lisbon | ? | | Brito |
| 1739 | | Demetrio | Schiassi | Lisbon | revival | 1732 | Brito |
| 1739 | | Siface | Leo | Lisbon | revival | 1737 | Corago |
| 1740 | | Alessandro | Fabbri | Lisbon | premiere | | Corago |
| 1740 | | Catone | Di Capua | Lisbon | premiere | | Brito |
| 1740 | | Ciro riconosciuto | Unknown | Lisbon | ? | | Brito |
| 1740 | | Ezio | Broschi | Lisbon | revival? | 1731? | Corago |
| 1741 | | Didone | Di Capua | Lisbon | premiere | | Brito |
| 1752 | 12 Sep | Siroe | Perez | Lisbon-CT | revival | 1740 | Brito |
| 1752 | 17 Dec | Demofoonte | Perez | Lisbon-CT | premiere | | Brito |
| 1753 | 21 Jan | Didone | Perez | Lisbon-CT | revival | 1751 | Brito |
| 1753 | 6 Jun | L'eroe cinese | Perez | Lisbon-CT | premiere | | Brito |
| 1753 | 31 Mar | Olimpiade | Perez | Lisbon-CT | premiere | | Brito |
| 1754 | Carniv. | Adriano | Perez | Lisbon | revival | 1752 | Brito |
| 1754 | 31 Mar | Ipermestra | Perez | Lisbon-CT | premiere | | Brito |
| 1754 | 6 Jun | Artaserse | Perez | Lisbon-CT | revival | 1748 | Brito |
| 1755 | 31 Mar | Alessandro | Perez | Lisbon-CT | premiere | | Jackson |
| 1755 | 6 Jun | Clemenza | Mazzoni | Lisbon-CT | premiere | | Hall |
| 1755 | 27 Oct | Antigono | Mazzoni | Lisbon-CT | rehearsal | | Hall |
| 1763 | | Temistocle | Unknown | Lisbon | ? | | Corago |
| 1765 | Carniv. | Demetrio | Perez | Lisbon | premiere | | Corago |
| 1765 | Summer | Didone | Perez, others | Lisbon | revival (pasticcio?) | 1751? | Brito |
| 1765 | Autumn | Semiramide | Perez | Lisbon | revival | 1749 | Corago |
| 1765 | Summer | Zenobia | Perez | Lisbon | revival | 1751 | Corago |
| 1766 | | Demetrio | Perez | Lisbon | revival | 1765 | Jackson |
| 1768 | Carniv. | Artaserse | Scolari | Lisbon | revival | 1757 | Corago |
| 1768 | Autumn | Demetrio | Perez | Lisbon | revival | 1765 | Corago |
| 1770 | 18 Jan | Il re pastore | Jommelli | Lisbon-CT | revival | 1764 | Brito |
| 1770 | 6 Jun | Nitteti | Jommelli | Lisbon-CT | revival | 1759 | Brito |
| 1771 | 6 Jun | Clemenza | Jommelli | Lisbon-CT | revival | 1765 | Corago |
| 1771 | Carniv. | Semiramide | Jommelli | Lisbon | premiere | | Corago |
| 1772 | Autumn | Antigono | Di Majo | Lisbon | revival | 1767 | Brito |
| 1772 | 6 Jun | Demofoonte | Perez | Porto | revival | 1752 | Brito |
| 1772 | 20 Apr | Ezio | Jommelli | Lisbon-CT | premiere | | Brito |
| 1774 | 31 Mar | Olimpiade | Jommelli | Lisbon-CT | revival | 1761 | Corago |
| 1774 | 6 Jun | Trionfo di Clelia | Jommelli | Lisbon-CT | premiere | | Brito |

| Year | Date | Opera | Composer(s) | City | Type | Orig. version | Ref. |
|---|---|---|---|---|---|---|---|
| 1775 | 6 Jun | *Demofoonte* | Jommelli | Lisbon-CT | revival | 1764 | Brito |
| 1776 | 6 Jun | *Alessandro* | Jommelli | Lisbon | revival | 1760 | Brito |
| 1798 | Summer | *Olimpiade* | Cimarosa | Lisbon | revival | 1784 | Corago |
| 1798 | Summer | *Didone* | Marino | Porto | premiere | | Corago |
| 1799 | 16 Oct | *Didone* | Marino | Lisbon | revival | 1798 | Corago |
| 1800 | | *Alessandro* | Caruso | Lisbon | revival | 1787 | US-Wc |
| 1801 | 1 Feb | *Artaserse* | Cimarosa | Lisbon | revival | 1784 | Corago |
| 1803 | 13 Dec | *Didone* | Various | Lisbon | pasticcio | | Corago |
| 1806 | Carniv. | *Alessandro* | Perez | Lisbon-CT | revival | 1755 | Jackson |
| 1806 | 18 Oct | *Artaserse* | Portugal | Lisbon-CT | premiere | | Marques |
| 1808 | 15 Aug | *Demofoonte* | Portugal | Lisbon-CT | premiere | | Marques |
| 1819 | | *Demofoonte* | Portugal | Lisbon | revival | 1808 | |
| 1821 | | *Clemenza* | Cimarosa | Lisbon | premiere | | Corago |

**Table 1b.** Performances of Metastasio's *drammi per musica* in Portugal, 1736-1821.

Besides the 16 versions without unknown author, 20 were *pasticcios*, 39 were premieres (20 in Spain and 19 in Portugal), and the remaining 101 (63.125% of the total) were revivals. Omitting productions by unknown composers, almost 60% of the *opera seria* heard in the Peninsula was specifically written for its theatres. Attesting to their adherence to international preferences, the composers that sounded in the Iberian venues were well-known figures, especially B. Galuppi (22), D. Perez (20), and N. Jommelli (18). J. A. Hasse's (4) lessened presence —none of his *opere serie* seems to have been performed in Portugal, and in Spain only until 1739— nonetheless stands out.[42]

During the 1730s, Iberian audiences enjoyed markedly contemporary music: the average difference between the dates of the premiere and the peninsular revivals is 3 years; see Figure 1. In the 1760s and the 1770s, the average difference reached a peak of ca. 10 years (in 1760-1764 specifically), increasing until circa 12 years around the following decades. This tendency is opposite to the number of operas premiered in the Iberian theatres (see Figure 2), derived from the dramatic suspension of operatic life in both countries after 1755 and 1758, respectively.[43] Again at the end of the 1770s, and especially in the 1780s, there seems to be a renewed interest in more contemporary music, this happening in the Madrid's court much later.[44]

---

[41] We do not include in Table 1b four productions (*Adriano*, 1752; *Demetrio*, 1753; *Zenobia*, 1754; and *Siroe*, 1756) listed in Jackson, "The operas of David Perez", pp. 32-37, but discarded in his article with Maurizio Dottori, "Perez, David [Davide]", *Grove Music Online*, <https://www.oxfordmusiconline.com/> [consulted: 17/03/2021], as well as by Brito, *Opera in Portugal*, and Pedro Miguel Gomes Januário, "Teatro Real de la Ópera del Tajo (1752-1755)", Ph.D. diss., Universidad Politécnica de Madrid, 2008. We also discard four productions (*Adriano*, 1752; *Demofoonte*, 1753; *L'eroe cinese*, 1754; and *Olimpiade*, 1754) listed in Aline Gallasch Hall, "A cenografia e a ópera em Portugal no século XVIII: os teatros régios: 1750-1793", Ph.D. diss., Universidade de Évora, 2012", p. 38, for which she does not provide reference; she actually mentions as her sources for her chronology Brito, *Opera in Portugal*, and Januário, "Teatro Real de la Ópera del Tajo", but none of them mentions these productions. The premiere of *Antigono* was planned for 4 November 1755, but was never premiered owing to the earthquake of 30 October. However, as indicated in Table 1b, on 27 October 1755, the king attended the rehearsal, as confirmed by the correspondence of the nuncio; see Hall, "A cenografia e a ópera em Portugal", pp. 107-108.

[42] Interestingly, several *intermezzi* by Hasse were performed in Madrid up to 1750, as confirmed by Teresa Casanova Sánchez de Vega, "Cinque buffi alla corte di Madrid. Fonti spagnole per lo studio dell'intermezzo, e suo repertorio, durante i regni di Filippo V e Ferdinando VI (1738-1758)", in *Entremets e intermezzi. Lo spettacolo nello spettacolo nel Rinascimento e nel Barocco*, ed. Gaetano Pitarresi (Reggio Calabria: Edizioni del Conservatorio di Musica "F. Cilea", 2020), pp. 255-274.

[43] The economic problems resulting from the Lisbon earthquake can also be traced in the scoring of the revived 1740s versions, which tended to be thinner than those for later settings, most likely to reduce the burden on the court's economy.

[44] See the colours in Figures 1 and 2 in the online version of this article in open access.

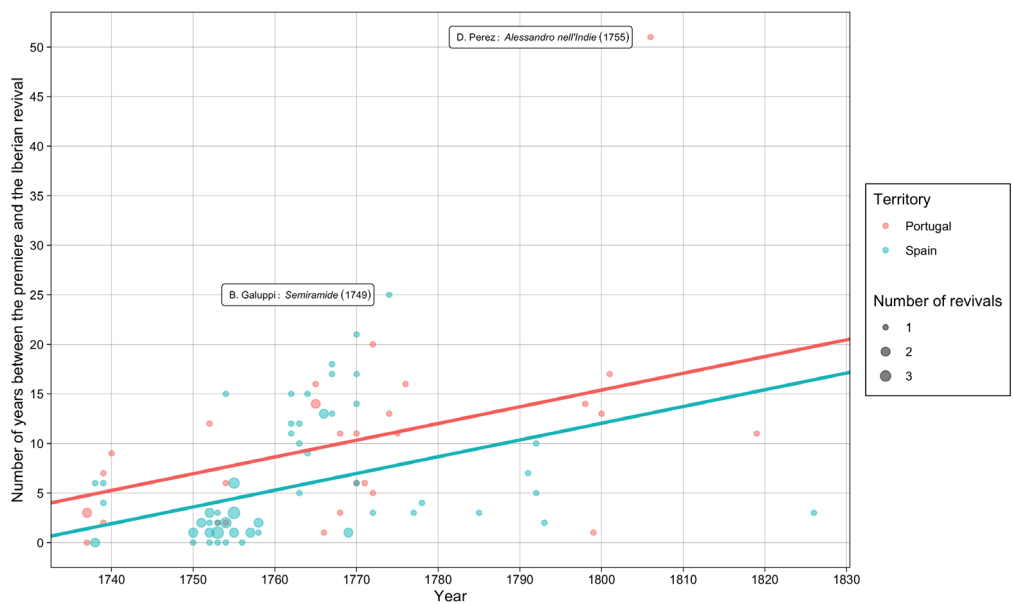

**Figure 1.** Metastasio's operas in the Iberian Peninsula, 1735-1775.

The solid lines represent the trends signaled by the optimal linear model obtained from a stepwise regression that accounts for the effects of the two territories. The fitted model indicates that the delay between the premiere and the Iberian revival increased an average of 0.17 years per each year increment in the considered time frame. It also indicates that the delay gap between Spain and Portugal remained temporally constant, with an estimated average delay of 3.35 years.

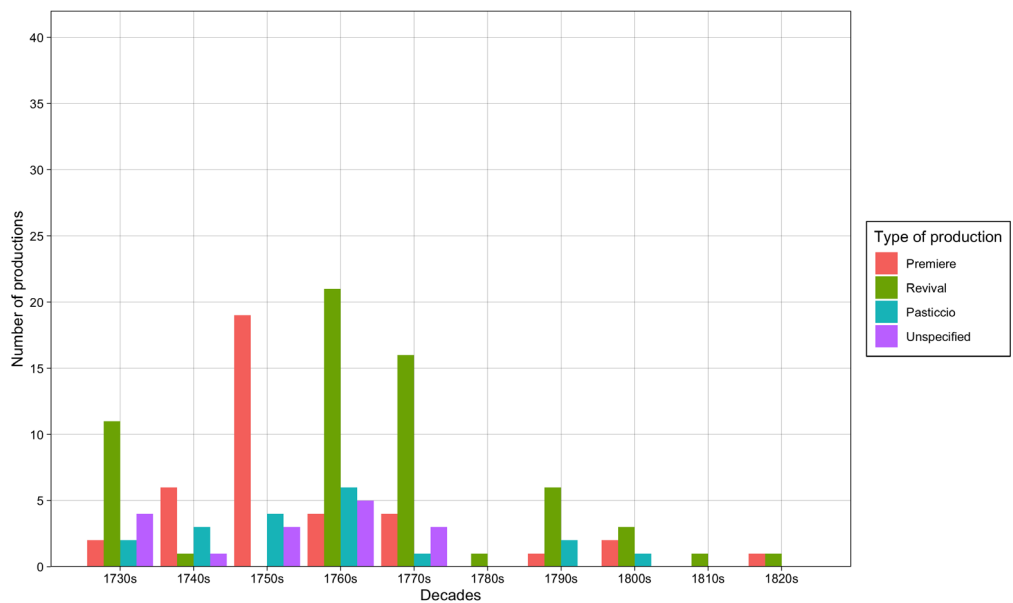

**Figure 2.** Total number of productions of Metastasio's *drammi* in the Iberian Peninsula, by decade.

The chronology and the repercussion of Metastasio's reception in the Peninsula seem clear. Yet "we still know very little about the specific modes of appropriation of Italian opera in […] a given court".[45] Metastasio's librettos tended in the Peninsula to be abridged for practical reasons,[46] i.e., to allow for easier courtly entertainment, "and not, as it regularly happened in the case of the commercial Italian opera, [to satisfy] the needs and impositions of the singers".[47] Queen Elisabeth Farnese herself supervised Corselli in the composition of arias for his *Alessandro nell'Indie*.[48] More influential, however, was queen Barbara of Braganza, whose appraisal, tested personally at the "cembalo reale", was decisive to approve or discard a score or even a composer for the court productions.[49]

In this context, several questions immediately arise: did not only the poets but also the European composers hired by the Peninsular rulers adapt their styles to satisfy the latter's tastes?[50] If so, how did those tastes materialise in music? Furthermore, if singers' demands were of little or no importance for the textual reshaping of Metastasio's poetry, were their usual technical requirements of little or no consequence to the composers' musical decisions too? In the specific case of Spain and perhaps also in Portugal,[51] did Spanish vernacular music, to which courtly audiences were accustomed, permeate opera composers' "international" writing?

To explore these issues, we have compiled a corpus of 2,404 arias from 126 musical versions of Metastasian operas composed in the eighteenth century,[52] both for the Iberian Peninsula and elsewhere in Europe; see the Appendix for a complete list.[53] For this corpus we discard revivals[54] and *pasticcios*,[55] in order to be able to ascertain compositional intentions. This also allows for a more consistent approach, as in the case of the Peninsula the scarcity of musical sources for the operatic productions makes it "not easy […] to know exactly which […] pieces heard in Spanish [and Portuguese] theatres correspond to those composed for the Italian operas in other European productions of [the same] titles".[56] In other words,

---

[45] Carreras, "En torno a la introducción de la ópera de corte", p. 325: "Sabemos todavía muy poco de las modalidades específicas de apropiación de la ópera italiana en […] una corte determinada".

[46] Gutiérrez Carou has studied the characteristics Metastasio's librettos abridged for Madrid. See Gutiérrez Carou, "The Textual Evolution of Metastasio's *Semiramide*" and Gutierrez Carou, "Evoluzione musicale e struttura letteraria".

[47] Carreras, "En torno a la introducción de la ópera de corte", p. 329: "… y no, como ocurría regularmente en el caso de la ópera comercial italiana, las necesidades e imposiciones de los cantantes".

[48] Carreras, "En torno a la introducción de la ópera de corte", p. 333.

[49] Farinelli's letters to Sicinio Pepoli are a clear witness to the queen's influence, particularly to despise Jommelli despite Farinelli's warm recommendation: "le composizioni sue non hanno avuto troppo incontro sopra a Cembalo Reale, dove si giudica da Sublime Persone". See Carlo Vitali, *Carlo Broschi Farinelli: la solitudine amica. Lettere al conte Sicinio Pepoli* (Palermo: Sellerio editore, 2000), quote dated 6 August 1749, on p. 188.

[50] The audience's preferences had little influence under Fernando VI, as court opera was the "King's private diversion", in the words of the British ambassador Benjamin Keene; see Juan José Carreras, "Farinelli's dream", pp. 357-393; quote on p. 373.

[51] Brito, *Opera in Portugal*, p. 2: "Spanish musical influence [was] still present, however, in the various comedies with music or in the *zarzuelas* which were performed in Lisbon until at least 1739".

[52] By decades, the number of arias in our corpus is: 1720s: 32; 1730s: 422; 1740s: 441; 1750s: 518; 1760s: 497; 1770s: 285; 1780s: 178; 1790s: 11. These ratios are in concordance with the number of Metastasian settings that were composed in the century, as can be tested by searching at <http://corago.unibo.it/>.

[53] Grouping together other European countries is just a methodological procedure to identify general tendencies beyond regional or national conventions yet it does not imply that their features were homogeneous. There were clear musical divergences depending on the audiences, as demonstrated by Strohm for Handel's London; see Reinhard Strohm, "L'*Alessandro nell'Indie* del Metastasio e le sue prime versioni musicali", in *La drammaturgia musicale*, ed. by Lorenzo Bianconi (Bologna: Il Mulino, 1998), pp. 157-175.

[54] Except in three cases in which no source for the premiere has been preserved: i) Perez's *Didone abbandonata*, which plausibly formed a new version. Arias in I-Vnm Mss.It. IV, 214-216 are therefore analysed and dated as corresponding to the 1753 Lisbonese version, not the 1751 Genovese premiere. ii) Traetta's *Didone abbandonata*. Arias in I-Nc Rari Cornice 5.28-30 are dated Milan 1763, instead of Venice 1758. iii) Sacchini's *Alessandro nell'Indie*. As a number of arias from the 1768 revival were not included in the premiere and as there is no record of Sacchini's participation in the Venetian revival, for this study only the arias present in both performances have been taken into account, being dated 1763.

[55] Yet versions originally by several composers are considered. See, for instance, Leo, Sarro, and Mancini's 1735 *Demofoonte*, among others.

[56] Leza, "El mestizaje ilustrado", p. 506: "Tampoco es fácil […] saber exactamente qué […] músicas escuchadas en los tea-

we only take into account versions for which authorship, date, and place of premiere are clear. Similarly, to minimise potential later modifications, we are using complete scores when available, although in the cases in which a complete version was composed but its source has not been preserved, we resort to incomplete versions of the manuscripts or for *arie sciolte*.[57]

More than 57% of the arias in our corpus were composed for Italian theatres, and 27% for venues in the Holy Roman Empire, this mostly comprising current Germany and Austria. Two were written for London, one for Denmark, and fourteen for the Iberian courts, which are the focus of this study:

— For Portugal (7): D. Perez's *Demofoonte* (1752), *Didone abbandonata* (1753), *L'eroe cinese* (1753), *Olimpiade* (1753), *Adriano in Siria* (1754), *Ipermestra* (1754), and *Alessandro nell'Indie* (1755).
— For Spain (7): F. Corselli's *Achille in Sciro* (1744), B. Galuppi's *Demofoonte* (1749); N. Jommelli's *Demetrio* (1751); B. Galuppi's *Didone abbandonata* (1752); and N. Conforto's *Siroe* (1752), *L'eroe cinese* (1754), and *Nitteti* (1756).

This amounts to a total of 89 arias purposefully written for Spain and 159 for Portugal. With a focus on the specific decades, it is important to note that 6.80% (30) of the 441 arias composed in the 1740s were crafted for the Spanish court, and that, among the 518 arias in the 1750s, 11.39% (59) were composed for Madrid and 27.80% (144) for Lisbon. This informs that the corpus is representative of the period's trends.

Before composing for the Iberian courts, all five composers had already enjoyed important careers elsewhere in Europe, especially Perez[58] and Galuppi.[59] Also Corselli and Conforto had already composed for Italian theatres[60] and Jommelli had been commissioned from theatres all over Europe,[61] all of this accounting for the Peninsular participation in the contemporary European operatic webs. This also applies to some of the singers involved in the fifteen premieres. For instance, Antonio Montagnana —Licomede in Corselli's *Achille*— had been a member of the cast in a number of operas composed by Händel, Porpora, Veracini, and Hasse.[62] Even more noteworthy are the cases of some performers taking part in the Lisbon premieres: Anton Raaf[63] —Iarba, Alessandro, Demofoonte, and Clisthene in Perez's *Didone, Alessandro, Demofoonte*, and *Olimpiade*—and Gaetano Majorano "Caffarelli" —Poro in his *Alessandro* of 1755. To ensure the validity of some of the subsequent analyses, we resort to data on other arias specifically written for these singers. In terms of numbers, Caffarelli affords the most balanced comparisons in the corpus studied here, as he also sang the role of Timante in the *Demofoonte* by Leo, Sarro, and Mancini, as well as that of Enea in both Jommelli's 1749 and Manna's 1751 settings of *Didone*.

Further attesting to Madrid and Lisbon's adherence to contemporary fashions, the fifteen select versions more-

---

tros españoles se corresponden con las compuestas para las óperas italianas en otras producciones europeas de estos títulos".

[57] This issue affects sources related to Madrid. Only the first two acts of Conforto's *L'eroe cinese* and the first act of Galuppi's *Didone* have survived. The other exception is Galuppi's *Demofoonte* (1749), for which we have 5 *arie sciolte* and Conforto's *Siroe* (1752), for which there are 2 *arie*. One duetto ("Vanne a regnar, ben mio") from Ferrandini's *Il re pastore* (1756) has been preserved, but for this study we consider arias only. Two arias from Mazzoni's *Il re pastore* (1757) have been preserved too ("Si spande al sole in faccia", D-Wa 6 Alt 694, and "L'amerò sarò costante", D-Hs M A/878 (Nr. 16)), but due to the COVID-19 pandemic we could not have access to them. See note in Table 1a regarding Ferrandini's *Semiramide* (1756).

[58] Mauricio Dottori and Paul J. Jackson, "Perez, David [Davide]", *Grove Music Online*, <https://www.oxfordmusiconline.com/> [consulted: 17/03/2021].

[59] Dale E. Monson, "Galuppi, Baldassare", *Grove Music Online*, <https://www.oxfordmusiconline.com/> [consulted: 17/03/2021].

[60] *La venere placata* for the Teatro San Samuele (1731) and *Nino* for the Teatro Sant'Angelo (1732). Álvaro Torrente, *Fiesta de Navidad en la Capilla Real de Madrid, V: Los villancicos de Francisco Corselli de 1743* (Madrid: Alpuerto, 2002), p. 18; Stiffoni, "Per una biografia del compositore napoletano Nicola Conforto"; and Robert Stevenson, "Conforto [Conforti], Nicola", *Grove Music Online*, <https://www.oxfordmusiconline.com/> [consulted: 17/03/2021].

[61] See, for example, Gaetano Pitarresi, ed., *Niccolò Jommelli: L'esperienza europea di un musicista 'filosofo'* (Reggio Calabria: Edizioni del Conservatorio di Musica "F. Cilea", 2014).

[62] Winton Dean, "Montagnana, Antonio Paolo", in *Grove Music Online*, <https://www.oxfordmusiconline.com/> [consulted: 17/03/2021]; and Giovanni Polin, "Montagnana, Antonio Paolo", in *Dizionario Biografico degli Italiani*, vol. 75 (2011) <https://www.treccani.it/> [consulted: 14/04/2021].

[63] On the relation between Raaf and the Spanish court, see Elisabetta Pasquini, "'Respinto da un impensato vento contrario in alto mare': Anton Raaff, il Farinelli e la *Storia della musica* di Giambattista Martini", *Recercare*, 29/1–2 (2017), pp. 181-252.

over concentrate in the years of Metastasio's heightened popularity in the rest of Europe, the 1740s and the 1750s.[64] Thus, before offering a broader picture, some of the ensuing analyses focus on the music composed in those two decades separately with a view to minimising the statistical impact that the general changes in musical style could have had on composers' writing throughout the century.

The analyses are organised by parameters of increasing specificity. First, we delve into high-level structural issues, such as form, metre, tempo, and key. As regards key, we explore it in detail too, observing trends apropos the key signature and the mode. Moreover, we focus not only on the composers' choices for the opening section of the —mostly tripartite— arias, but also on how they devised the metric, tonal, and tempo changes —if any— for the middle sections. To close the study, we turn to the composers' treatment of the singers' voices, with a special focus on the compass for the various voice types and for singers for which such and other composers wrote arias elsewhere in Europe. As we have discussed above, the conception of the operatic productions too was markedly different in the 1740s and the 1750s. Therefore, besides offering an overview of the central years of the eighteenth century, we also provide insights into the two decades separately, were there to be any significant divergences.

The majority of the exploratory analyses below are mainly based on the comparison of the proportions of specific traits in the arias corresponding to the three territories for which we have sources for the 1740s and the 1750s, i.e., Spain, Portugal, and central Europe and Italy. Yet, the comparison of such proportions does not directly inform on the existence of differences between territories that are unlikely to be produced either by the corpus' inherent uncertainty or by the effect of other factors that are unaccounted for in the analysis. Of particular importance is the fact that all the operas for the Portuguese court analysed here were composed by the same person, David Perez, while among the 30 arias for Madrid in the 1740s, 25 come from *Achille in Sciro* by Corselli.

Thus, to formally validate —or discard— the conclusions drawn from such analyses, we have performed two hypothesis tests that are customary in statistics. The first of these is the Chi-Squared Test of Independence for contingency tables, henceforth CSTI, which has helped us to determine whether certain traits are homogenously distributed across territories or, on the contrary, whether there exist significant differences that are unlikely to result from chance or from composers' individual preferences. The second of the tests, the one-sided two-sample Test of Proportions, henceforth TP, has been performed to evaluate whether the distribution of specific traits among several territories can be regarded as a territorial peculiarity. Preliminarily, the probability of a single "false discovery" (i.e., incorrectly flagging as false a null hypothesis that is true) in these tests corresponds to α = 0.05, i.e., results below 5% indicate statistical significance. However, to further validate our claims, a conservative Bonferroni Correction Factor[65] has been applied to all tests when assessing their outcomes to account for the number of tests performed (N = 26), as the higher the number of tests, the higher the probability of obtaining a single statistically significant result merely by chance. In this manner, we control the probability of false discoveries so that only $p$-values below α/N, i.e., $p \leq 0.0019$, are considered as statistically significant. In any case, the discussion of results that both are and are not statistically significant will offer interesting insights into the stylistic differences among the three select territories.

## ARIA FORM

Musicology has tended to believe that the formal changes in operatic arias throughout the eighteenth century —with a progressive abandonment of the *da capo* mould for the *dal segno* and, later, binary or other forms— were a consequence of general, pan-European practices related to balance and dramatic verisimilitude.[66]

---

[64] See note 48 above.

[65] For a technical explanation of the two tests and the Bonferroni Correction Factor, see, for instance, David J. Sheskin, *Handbook of Parametric and Nonparametric Statistical Procedures*, 5th ed. (London: Chapman and Hall/CRC, 2011).

[66] Several studies support this thesis. See Michael H. Arshagouni, "Aria forms in opera seria of the classic period: Settings of Metastasio's *Artaserse* from 1760–1790", Ph.D. diss., University of California, 1994; Mary K. Hunter, "Haydn's aria forms: A study of the arias in the Italian operas written at Eszterhaza, 1766–1790", Ph.D. diss., Cornell University, 1982; Charles Rosen, "Aria", in *Sonata Forms*, rev. ed. (New York and London: W. W. Norton, 1988), pp. 28-70; Andrea Chegai, "Forme limite ed eccezioni formali in mezzo secolo di intonazioni metastasiane. Cavatine, arie pluristrofiche, rondò e altro", in *Il canto di Mestastasio*, ed. Maria Giovanna Miggiani (Venice: Arnaldo Forni Editore, 2004), pp. 341-408; and Paul M. Sherrill, "The Metastasian Da Capo aria: Moral philosophy, characteristic actions, and dialogic form", Ph.D. diss., Indiana University, 2016. See also the articles in Lorenzo Bianconi and Michel Noray, eds., *L'aria col da capo* (Bologna: Il Mulino, 2008).

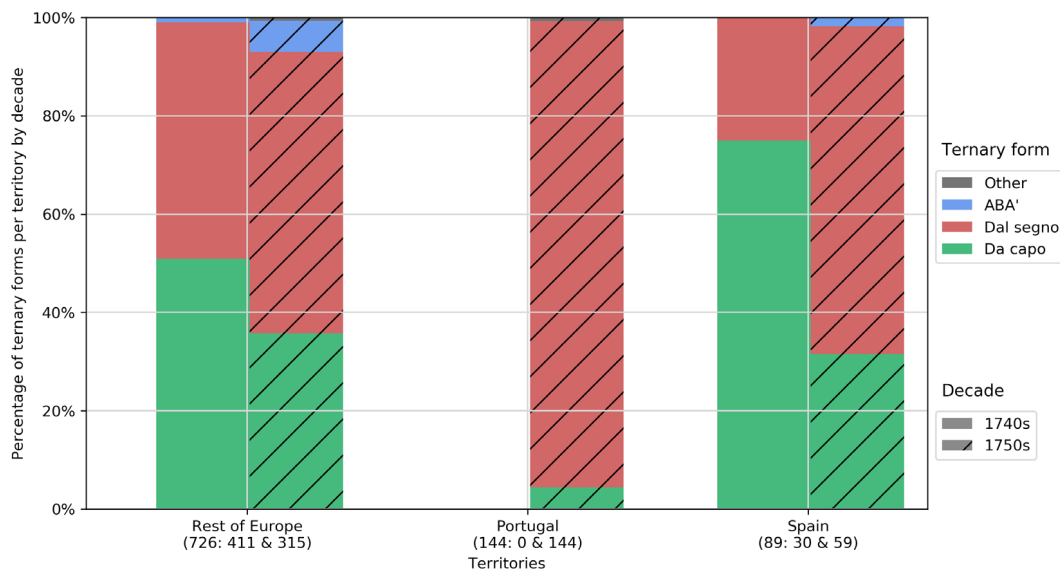

**Figure 3.** Ternary forms in Metastasian arias, 1740s and 1750s: differences among the various territories. The stacked bars represent the proportion of the various formal types among the ternary forms.

However, the statistical observation of a corpus as large and varied as ours opens new practical and theoretical scenarios.

Unsurprisingly, the corpus in the 1740s and the 1750s is dominated by ternary formal types (*da capo*, *dal segno*, ABA', and, rarely, ABC), as dictated by Metastasio's poetic texts.[67] The ratio of ternary forms is in the Peninsula similar to that in the rest of Europe (ca. 95%). However, if we restrict the analysis to ternary forms only, territorial differences become more telling, especially if we observe the two decades separately. In the first of them, composers to the Spanish court seem to have stuck to the more traditional *da capo* type (75%), while the tendency in the rest of Europe was already to move away towards the *dal segno* construction (48.11%); in fact, that would be the predominant ternary mould in Europe in the 1760s, with 76.33% of the arias.

In any case, in the 1750s, according to the available data, the scenario in Madrid appears to change:

whereas in Corselli's *Achille* (1744) and Galuppi's *Demofoonte* (1749) 75% of the arias are of the *da capo* type, later on, with Jommelli and Conforto, as well as with Galuppi's *Didone* (1752), it is the *dal segno* type that features most prominently (66.66%); see Figure 3. In Portugal, Perez set Metastasio's texts almost exclusively to the *dal segno* form (94.89% of the ternary arias premiered in Lisbon), which brings him closer to 1760s proportions (76.33% in the whole corpus). Interestingly, in his *Artaserse* for Naples in 1749, three years before the premiere of his Portuguese *Demofoonte*, only 60.86% of the ternary arias are of the *dal segno* type, whereas the remaining 39% are set to a *da capo* schema. All in all, this points to a broad change in paradigm around the middle of the century, a change in which Portugal, or David Perez more exactly, was ahead most of his contemporaries.[68]

---

[67] One important exception to this is "Se cerca, se dice" from his *Olimpiade*, which, by editorial tradition, soon became a binary form. For a detailed study of this tradition, both poetic and musical, see Nathaniel D. Mitchell, "The 'Se cerca' script: Conventions and creativity in an eighteenth-century aria tradition", Ph.D. diss., Princeton University, 2020, esp. pp. 170-181.

[68] A CSTI on the 1750s data returns $p = 3.354 \cdot 10\text{-}12$, indicating that the differences among the do not result from chance. On the contrary, a $p = 0.045$ for the 1740s data suggests that personal preferences may have played a part. The TP confirms the hypothesis that Perez's use of *dal segno* forms in the 1750s was significantly more marked than the practice both in Spain and in the rest of Europe, with $p = 0$ in both cases.

| Metre | | | | |
|---|---|---|---|---|
| **Decade** | **Simple duple** | **Simple triple** | **Compound duple** | **Compound triple** |
| 1720s | 71.88% | 25.00% | 3.13% | 0.00% |
| 1730s | 74.64% | 23.46% | 1.90% | 0.00% |
| 1740s | 68.25% | 28.79% | 2.95% | 0.00% |
| 1750s | 70.66% | 27.02% | 2.02% | 0.002% |
| 1760s | 75.45% | 23.34% | 1.20% | 0.00% |
| 1770s | 76.84% | 21.75% | 1.40% | 0.00% |
| 1780s | 78.09% | 20.22% | 1.69% | 0.00% |
| 1790s | 100.00% | 0.00% | 0.00% | 0.00% |

**Table 2.** Metre in Metastasian arias throughout the eighteenth century, by decade.

## TIME SIGNATURE

One may hypothesise that composers' choice of specific, or at least types of, time signatures was dictated by the accentuation and the length of Metastasio's verses.[69] Yet, whereas some specific cases, such as the settings of *Didone abbandonata*'s "Son regina" (87% are in simple duple time),[70] support such assumption, most of the data point in a different direction. "Se mai turbo il tuo riposo" and "Se mai più sarò geloso" from *Alessandro nell'Indie* form an interesting case in point. The cavatina and the aria merge in the first act into a duet for the main lovers in the plot, Poro and Cleofide. One would expect that composers would set the two pieces to the same metres so the fusion in the *duetto* would be more straightforward. Yet that is not the case. Not only do the settings diverge apropos the time signatures (the most frequent is C, yet it does not appear in more than 34% of the pieces) and even the types thereof (more than 25% of the pieces are in simple triple time, with the rest being in simple duple): 6 (30%) out of the 20 *Alessandros* in the corpus present different metric types for Poro's cavatina and Cleofide's aria and further 8 (40%) have a different time signature although of the same metric type.

An even more striking case is *Demofoonte*'s "Misero pargoletto". 13 (60%) of the 22 such arias that we have analysed are in simple duple time, 8 in simple triple, and 1 in compound duple. Among those in simple duple, there is no consensus either: while the majority are set to C|, there are examples of 2/4 and C too. Although one may refute the analysis by arguing that the corpus by no means comprises all the "Misero pargoletto" settings that were composed in the eighteenth century, our conclusions are supported by a broader study. According to it, the settings "divide evenly between simple duple/quadruple and simple triple times", with the 3% being in compound duple time.[71] This further validates our corpus as significant and reliable among Metastasian operas.

In the use of metre, then, one needs to seek for explanations other than the poetic structure. Interestingly, there appears to be an increasing preference towards binary metres and, more specifically, towards simple duple time as the second half of the eighteenth century progresses, with an accompanying decrease in the frequency of arias in ternary time signatures. The use of 9/8 is in our corpus still very scarce, with only 1 example in the 1750s; see Table 2.

The arias for Spain show a higher preference towards triple time in the 1740s and a marked increase in the ratio of simple duple time in the 1750s. Whereas the presence of this latter metre in the 1740s may seem to be significantly lower than in the rest of Europe (56.66% vs 69.09%), the TP for such a hypothesis returns a $p = 0.079$, which indicates that the differences are not large enough to be considered as a Spanish peculiarity. Perhaps, the differences lie more on the use —or avoidance— of specific time signatures. For instance, in the Spanish settings there is no single aria in 6/8, 12/8, or 9/8, and also a proportionally larger number of arias in C| were heard; see Figure 4.

---

[69] For further discussion, see N. D. Mitchell, "The 'Se cerca' script", esp. p. 151.

[70] We classify relevant time signatures as follows: i) simple duple time: 2/4, 4/4, C, C|; ii) simple triple time: 3/8, 3/4, 3/2; iii) compound duple time: 6/8, 12/8; iv) compound triple time: 9/8.

[71] Álvaro Torrente and Ana Llorens, "'Misero pargoletto': Kinship, Taboo and Passion in Metastasio's *Demofoonte*", in *Demofoonte come soggetto per il dramma per musica: Johann Adolph Hasse ed altri compositori del Settecento*, ed. Milada Jonášová and Tomislav Volek (Prague: Academia, 2020), p. 81.

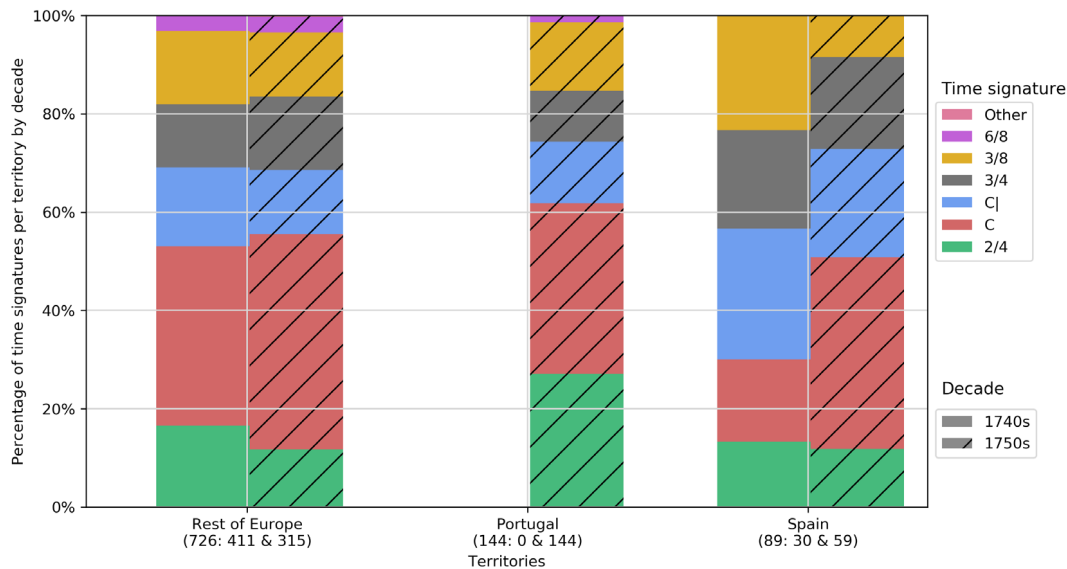

**Figure 4.** Main time signatures in Metastasian arias, 1740s and 1750s: differences among the various territories. The stacked bars represent the proportion of the various signatures for the opening sections in the arias.

Although the majority use of binary time signatures may plausibly indicate that Spain was in line with the general practices of the time, in reality the use of cut time ( C|) in Spain was characteristically greater in the 1740s ($p = 0.067$ in the TP with respect to the rest of Europe) and, specially, in the 1750s ($p = 0.028$ including Portugal in the comparison). This may have been in relation to the drastic, yet consistent, change that the metric features of Spanish vernacular —and to a certain extent also Italian— music had suffered in the preceding two centuries. Such music in Spain shows a marked preference for binary meters during the central part of the sixteenth century while a radical shift towards ternary meters took place from the 1580s in coincidence with the rise of *romance nuevo*.[72] Seventeenth-century Italian opera also reveals a preference for ternary meter in the arias, particularly in the works of Francesco Cavalli. A new metric revolution took place in Spain the early decades of the eighteenth century implying the departure from mensural notation into a full ortochronic system, with an increasing preference for binary meters in music.[73] The still preliminary study of vernacular music in public theatres —with a marked preference for ternary rhythms such as *seguidillas*— reveals a vivid contrast with the increasing use of binary meter in operatic arias confirmed in our corpus. Perhaps for this reason composers in the enlightened circles of the court took pains to detach themselves from the popular aesthetics of the immediately previous period that were still heard in the *corrales*. And they did so through the comparatively pre-eminent use of the *alla breve* time signature and not by resorting to or avoiding specific ternary time signatures.[74] This may therefore have been a means to differentiate court music from native musical traditions, where C| was not at all frequent.[75] Perhaps the *alla breve* was composers' way of providing the music with a unique, courtly flavour that was nonetheless not too alien to the general practices in the rest of Europe.

---

[72] As analysed in Álvaro Torrente, "Tonos, bailes y guitarras: la música en los ámbitos privados", in *Historia de la música en España e Hispanoamérica*, vol. 3: *La música en el siglo XVII*, ed. Álvaro Torrente (Madrid: Fondo de Cultura Económica, 2016), esp. pp. 199-202.

[73] See Álvaro Torrente, "La modernización/italianización de la música sacra", in *Historia de la música en España e Hispanoamérica*, vol. 4: *La música en el siglo XVIII*, ed. José Máximo Leza (Madrid: Fondo de Cultura Económica, 2015), pp. 125-156.

[74] Supporting this claim, the apparent preeminence of 3/4 in Spain in the 1740s is not stylistically significant, with $p = 0.135$ in the TP in the comparison with the rest of Europe. Also the ratio of arias in 3/8 is not notably lesser either ($p = 0.107$).

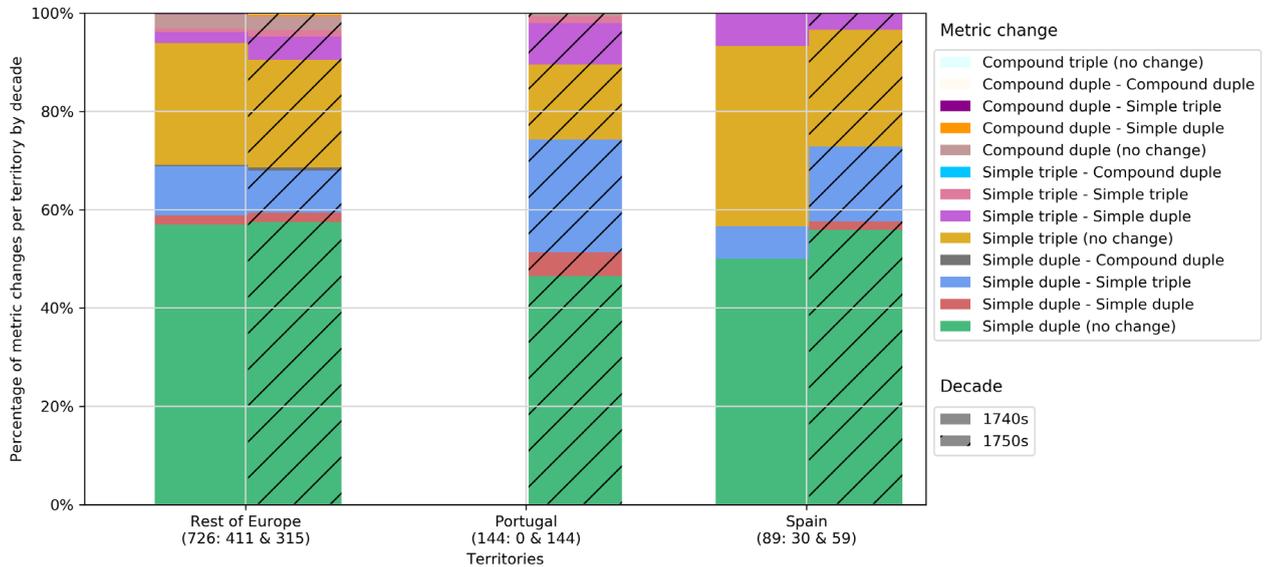

**Figure 5.** Change of metre between the A and the B sections in Metastasian arias, 1740s and 1750s: differences among the various territories. The stacked bars represent the proportion of the various metric contrasts.

Among the non-pervasive metric contrasts between the A and the B sections of the arias, composers for the Bourbon court in the 1740s —mostly Corselli— invariably resorted to the simplest types, i.e., from simple duple to simple triple time and vice versa; in fact, they would never move towards a different time signature of the same type or use a compound metric type in either the A or B section. For their part, composers writing for theatres in the rest of the continent infused their arias with greater metric variety, except for the change from simple triple to compound duple time; see Figure 5.[76] This greater variety in central Europe and Italy would become more pronounced in the 1750s, although we cannot speak of an indisputable territorial differentiation in this regard (in the CSTI, $p = 0.566$ on the 1740s and $p = 0.007$ on the 1750s data). As in the case of the specific time signatures, the differentiation may rest on the use of specific metric changes for the B sections. Specifically, in both Spain and Portugal, still in the 1750s the change from simple duple to simple triple time was the most commonly employed (15.25% and 22.92% respectively, vs. 8.57% in the rest of the continent), this becoming a significant compositional trait of the Peninsular arias of the 1750s ($p = 0$ in the TP).

## TEMPO

There seems to be a general tendency towards moderate tempi[77] across the middle of the century (Figure 6a), especially in Spain and Portugal (45.76% of the arias for Spain and 42.36% of the arias for Portugal vs. 34.29% in the rest of Europe in the 1750s; Figure 6b), yet the data do not wholly ascertain a trend specific to any of the three territories.[78] However, one can observe a certain Peninsular preference for non-slow tempi, i.e., for tempi that can be classified as fast or moderate ($p = 0.010$ in the TP when comparing the Peninsular arias with those for elsewhere in Europe).

---

[75] For a study of metre in Spanish vernacular music, see Álvaro Torrente, "Sacred villancico in early eighteenth-century Spain: The repertory of Salamanca Cathedral", Ph.D. diss., University of Cambridge, 1998, esp. p. 141, as well as his "Tonos, bailes y guitarras: la música en los ámbitos privados", in *Historia de la música en España e Hispanoamérica*, vol. 3: *La música en el siglo XVII*, ed. Álvaro Torrente (Madrid: Fondo de Cultura Económica, 2016), pp. 199-202.

[76] Interestingly, the only aria in compound triple time, Hasse's "Consola il genitore" (A3-S7) from his Dresden *Olimpiade* (1756), remains in 9/8 for the B section.

[77] We classify the tempi as follows: i) slow tempi: Adagietto, Adagio, Grave, Larghetto, Largo, Lento, Sostenuto; and combinations with these; ii) moderate tempi: A tempo, Affettuoso, Amoroso, Andante, Andantino, Arioso, Cantabile, Comodo, Espressivo, Grazioso, Gustuso, Maestoso, Minuetto, Moderato, Parlante; and combinations with these; iii) fast tempi: Agitato, Allegretto, Allegro, Con bravura, Con brio, Con spirito, Furioso, Presto, Risoluto, Spiritoso, Vivace; and combinations with these.

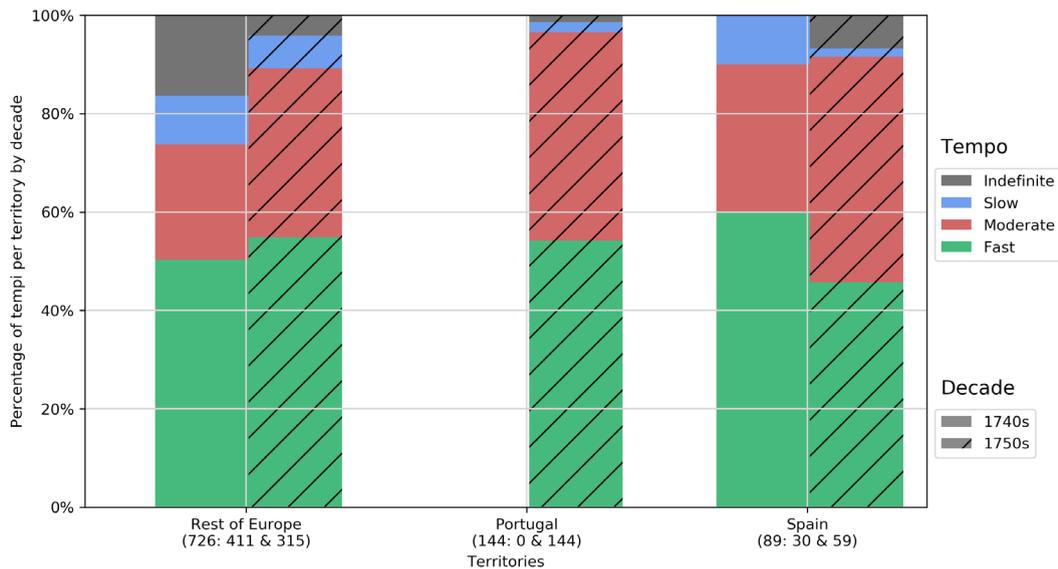

**Figure 6a.** Main tempo in Metastasian arias, 1740s and 1750s: differences among the various territories. The stacked bars represent the proportion of the various tempo classes, as indicated in the opening sections in the arias.

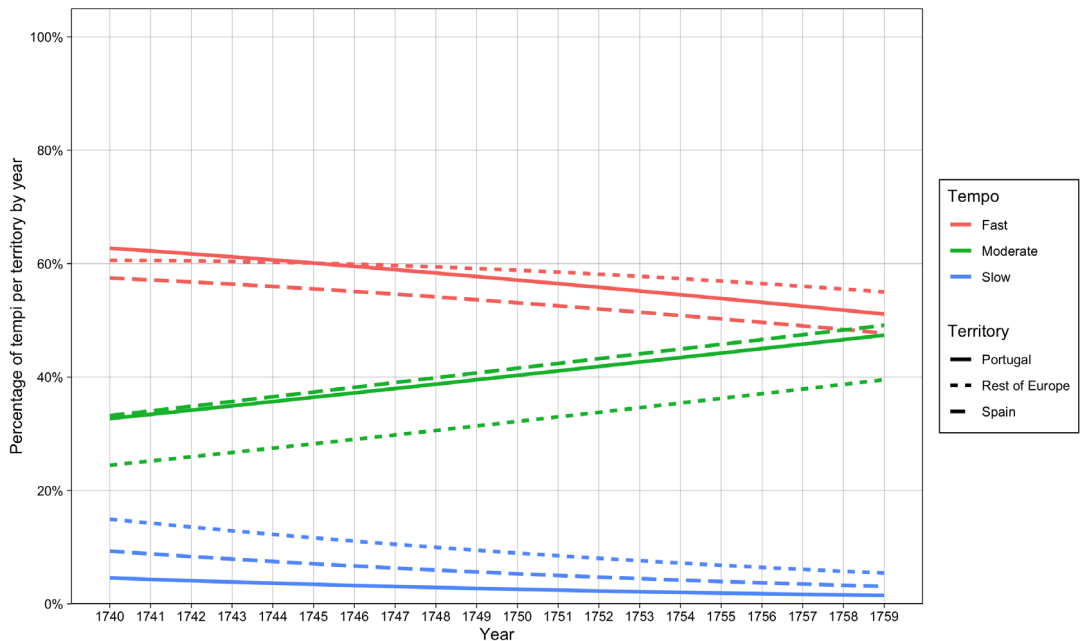

**Figure 6b.** Increasing preponderance of moderate tempi, 1740-1749. The curves represent the proportion trend of each tempo class, as estimated from a multinomial regression model featuring the year and the territory as predictors, and the tempo as response.

|  |  | 1740s | | | 1750s | | |
|---|---|---|---|---|---|---|---|
|  |  | Fast | Moderate | Slow | Fast | Moderate | Slow |
| **Rest of Europe** | Fast | 9 | 8 | 4 | 3 | 4 | 6 |
|  | Moderate | 12 | 7 | 3 | 10 | 12 | 0 |
|  | Slow | 10 | 4 | 2 | 6 | 3 | 1 |
| **Portugal** | Fast |  |  |  | 1 | 14 | 0 |
|  | Moderate |  |  |  | 10 | 24 | 1 |
|  | Slow |  |  |  | 1 | 1 | 0 |
| **Spain** | Fast | 0 | 0 | 0 | 3 | 1 | 1 |
|  | Moderate | 1 | 0 | 0 | 3 | 5 | 0 |
|  | Slow | 1 | 1 | 0 | 0 | 0 | 0 |

**Table 3.** Metastasian arias with a change of tempo for the B section, 1740s and 1750s: differences among the various territories.

Again, the changes between the A and the B sections open a richer picture. It is striking that, even if the ratio of tempo change increases in the 1750s with respect to the 1740s, in the two decades as a whole in the Peninsular arias there are significantly more changes of tempo than in the arias written for elsewhere in Europe (31.05% vs. 16.39%, null *p*-value for the TP).[79] As expected, changes departing from fast or moderate tempi are more frequent in the Peninsular operas dating from the 1750s; see Table 3. In Spain there is moreover a marked tendency to mild changes between fast tempi in both sections (5.08% vs 0.97% in the rest of Europe and 1% in Portugal), and, in fact, data lead us to affirm a territorial trend ($p = 0.001$ when applying a TP to the data on Spain on the one hand with Portugal and the rest of Europe on the other).

Whereas the percentage of arias in which the B section presents a change of tempo and of metre was similar among Spain and the rest of Europe (ca. 10% in the 1740s and ca. 15% in the 1750s), Perez used many more such marked contrasts (35.42%), pointing towards a clear differentiation (in TP, $p = 0$). More specifically, the number of different conjunct changes of tempo and metre in the B sections was in Spain much more reduced than elsewhere, including Portugal, in both the 1740s and the 1750. However, the statistical evaluation of this feature does not allow us to propose this a stylistic trait (CSTI: $p = 0.969$ in the 1740s and $p = 0.471$ in the 1750s), as this depends on the number of arias with such changes, which, in absolute terms, is reduced in the corpus, especially for Spain (3 for the 1740s and 11 for the 1750s).

**KEY**

The study of keys is more complex, as several factors may have played a role, such as the scoring, the pitch standards, and the tuning systems, and even the particular singers for which the arias were written. Our corpus as a whole shows a tendency towards fewer minor-mode arias as the decades elapse.[80] Yet the variation among decades

---

[78] TP when evaluating Spain and Portugal jointly with respect to the rest of Europe: $p = 0.113$.

[79] A similar divergence is supported by the results of the CSTI on the three samples ($p = 8.652 \cdot 10^{-9}$).

[80] The minor mode was used with decreasing frequency throughout the eighteenth century; see Rey M. Longyear, "The minor mode in eighteenth-century sonata form", *Journal of Music Theory*, 12/1-2 (1971), p. 182. The analysis of 5,750 eighteenth-century compositions revealed a tendency similar to that observed in our corpus: there is a decrease in the minor mode from the start of the eighteenth century (55.87%) up to the 1760s (8.33%) —in our case the 1790s—, with a subsequent increase until the end of the century (20%). In sum, the ratio of minor-mode compositions in both corpora is at the end of the eighteenth century lesser than at its beginning. For an analysis of the dataset, see Daniel Harasim, Fabian C. Moss, Matthias Ramirez, and Martin Rohrmeier, "Exploring the foundations of tonality: Statistical cognitive modeling of modes in the history of Western classical music", *Humanities and Social Sciences*

is slight and, thus, we do not analyse the 1740s and the 1750s separately. In general (see Figure 7), Perez seems to have frequently used F major and E-flat major, pointing towards a personal preference for the "soft"[81] and "grave"[82] flats among the dominating major-mode keys.[83] Also, among the minor keys specifically (see Figures 8a and 8b), he used the "pathetic" and "lamenting"[84] f minor significantly more often (75% of the minor-mode arias) than composers writing for theatres in the rest of Europe (17.18%).[85]

Although this may be in relation to specific tuning systems and transposition practices for the winds, it may perhaps attest to a more deeply rooted preference too. Tellingly, in his Portuguese versions Perez avoided the "gentle" C major[86] as well as A major,[87] indicating a significant reluctance to bright —i.e., on the sharp side of the circle of fifths— major-mode tonalities,[88] parallel to his exclusion of minor-mode sharp and natural keys: none of his arias are in a minor, e minor, b minor, or f-sharp minor.[89] All of this could potentially characterise Perez as tending towards sad aria settings when writing for the Portuguese court.

Even more interesting is the tonal relation between the keys in the two main sections of binary and ternary arias. As expected, most of the arias present a change towards the relative key (average of 52.61% in the whole corpus for the 1740s and the 1750s). Then, overall, the second most frequent change is towards the major subdominant key (21.77%), followed by the parallel tonic (6.77%) and other less-represented tonal relations. However, when observing the three select territories and the two decades separately (see Figure 9), meaningful divergences emerge especially in the 1740s (CSTI: $p = 2.496 \cdot 10^{-9}$ for the 1740s and $p = 0.002$ for the 1750s). In the 1750s, Perez seems to depart from the rule and play more with subdominant relations (33.33%, TP: $p = 0.001$ when comparing Portugal with the rest of Europe, in this case including Spain), which would enjoy outmost popularity (29.52%) in the following decade. At the same time, his central sections would be set in the relative key not as often (35.41%) as in the rest of Europe ($p = 0$).

---

*Communications*, 8 (2021), pp. 1-11. The dataset itself is available at <https://github.com/DCMLab/HistoryModes_DataCode/>.

[81] Jean-Philippe Rameau, *Observations sur notre instinct por la musique* (Paris: Praut fils, 1754), p. 54, as quoted and translated in Rita Steblin, *A History of Key Characteristics in the Eighteenth and Early Nineteenth Centuries*, 2nd ed. (Rochester: University of Rochester Press, 2002), p. 97.

[82] Rousseau (ca. 1749), Lacombe (1758), and Grétry (1797) associated the flat majors with majesty and gravity, and the sharp majors with "brilliance"; "…similarly, all three assign a 'touching' or 'pathetic' affect to the flat minors. No such correspondence exists […] a century earlier", attesting to the emergence of a sharp-flat principle in key characteristics. See Steblin, *A History of Key Characteristics,* p. 103.

[83] F major: 16.91% in Portugal, 13.75% in Spain, and 14.35% in the rest of Europe (in the TP, $p = 0.186$ when comparing Portugal with the other territories jointly); Eb major: 11.76% in Portugal, 6.65% in Spain, and 8.15% in the rest of Europe ($p = 0.006$). In this context, his notable use of E major (11%) is striking ($p = 0.003$ in TP).

[84] Gioseffo Zarlino, *Le Institutioni harmoniche* (1558), as quoted in Steblin, *A History of Key Characteristics*, p. 28.

[85] The *p*-value from the TP on the ratio of f-minor arias in Portugal on the one hand and the rest of Europe, including Spain, on the other is 0, which further attests to territorial preferences apropos this particular key among minor-mode tonalities.

[86] Jean Rousseau, *Traité de l'harmonie réduite à ses príncipes naturels* (Paris: Christophe Ballard, 1722), as quoted in Steblin, *A History of Key Characteristics*, p. 39, Table 3.2.

[87] C major: 4.41% in Portugal, 10.42% in Spain, and 11.63% in the rest of Europe; A major: 5.14% in Portugal, 13.74% in Spain, 13.14% in the rest of Europe; $p = 0$ when jointly comparing the use of the two keys in Portugal and the two other territories among major-mode arias.

[88] In his *Artaserse* for Naples (1749), however, 2 of the 24 arias are in C major, 4 are in G major, another 4 in D major, 3 in A major, and 2 in E major, supporting the claim that his avoidance of such keys in the Portuguese settings was a trait peculiar to them. The remaining arias in this *Artaserse* are in Bb major (5), F major (2), c minor (1), and f minor (1).

[89] This mostly explains the darkness of Perez's version, as observed by Lorenzo Bianconi, "Le 'mutazioni sceniche' nel teatro d'opera: immagini organizzate nel tempo", in *I Bibiena, una famiglia europea*, ed. by Deanna Lenzi and Jadranka Bentini (Venice: Marsilio, 2000), pp. 69-74. As Steblin notes, several factors "helped to associate the sad and languid affect traditionally ascribed to the minor mode with the flat, and the lively and cheerful affect of the major mode with the sharp"; see Steblin, *A History of Key Characteristics*, p. 96) Also, Rousseau spoke of "the flat minor keys"; see Jean Rousseau, *Méthode claire*, 4th ed. (Paris: Christophe Ballard, 1691), p. 15, as quoted in Steblin, *A History of Key Characteristics*, p. 99.

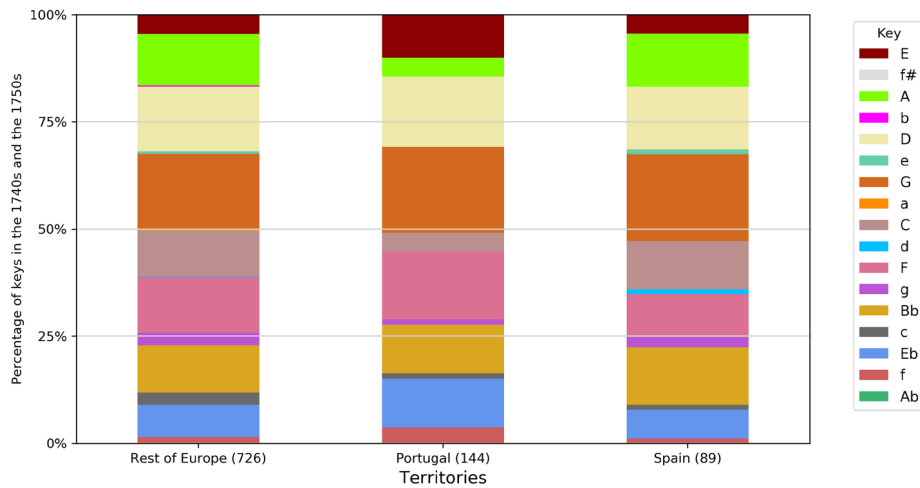

**Figure 7.** Ratio of keys in Metastasian arias, 1740s-1750s: differences among the various territories.

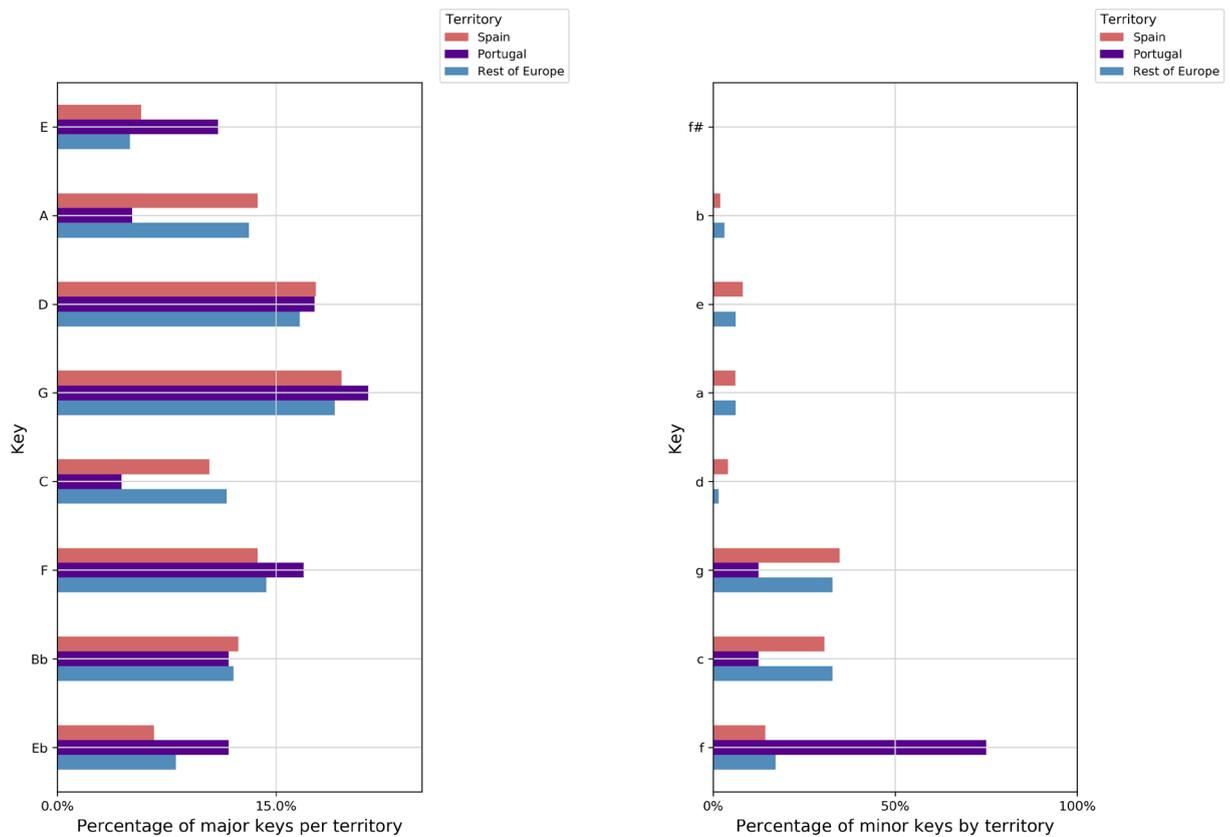

**Figures 8a and 8b**. Ratio of keys among their respective modes (major and minor) in Metastasian arias, 1740s-1750s: differences among the various territories.

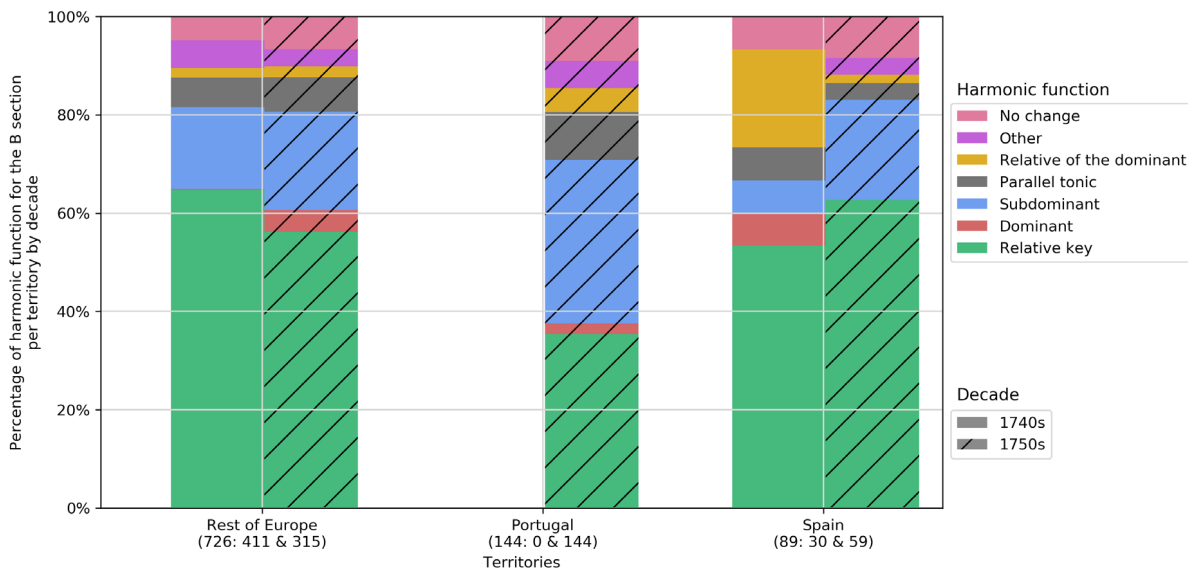

**Figure 9.** Percentage of harmonic function for the B section per territory by decade.

Arias from the Spanish court in the 1740s would contrarily opt for emphasising the dominant key, either directly (6.66%) or through its minor relative (20%) (TP: $p = 0$). Curiously, those ratios would progressively —yet softly— increase in our corpus as the century elapses. A plausible influence in this regard could be the French-style dance forms, fully integrated into Felipe V's court structure,[90] in which the B sections would normally start in the dominant key,[91] or even Domenico Scarlatti's binary keyboard sonatas.[92] This could be related to the fact that Francesco Corselli, the main composer in Spain in the 1740, was the son of the dance master to the Farnese court in Parma.

**VOCAL TREATMENT**

Significant as these divergences might be, the most provocative ones relate to the treatment of the vocal part.

In the corpus, bass voices are most often used in Central Europe, England, and Spain, while in Italy it was only Vivaldi who used them for one aria in his Venetian *Olimpiade* of 1734.[93] This may point not only to an Italian and Portuguese preference for higher voices but also to a wider circulation of musicians from/to Spain.

Such bass voices (1.08% of the corpus)[94] were treated very differently in Spain on the one hand and in the rest of Europe (Vienna, the Hubertusburg theatre, Dresden, Copenhagen, Bratislava, Venice, Naples, and London) on the other. In Spain, very conspicuously, Corselli exploit-ed the lowest range of Antonio Montagnana's compass, reaching a F2[95] and even E2. Similarly, the highest part of his compass was very little exploited, whereas in other versions basses had to reach even a F4.

---

[90] Pilar Montoya, "La recepción del estilo francés en los tratados de danza españoles del siglo XVIII", *Cuadernos dieciochistas*, 16 (2018), p. 40.

[91] For a survey of, for instance, the minuet form, see Meredith Ellis Little, "Minuet (Fr. menuet; Ger. Menuett; It. minuetto; Sp. minuete, minué)", *Grove Music Online*, <https://www.oxfordmusiconline.com/> [consulted: 23/03/2021].

[92] W. Dean Sutcliffe, "Formal dynamic", in *The Keyboard Sonatas of Domenico Scarlatti and Eighteenth-Century Musical Style* (Cambridge: Cambridge University Press, 2003), esp. p. 340.

[93] "Qual serpe tortuosa" (A2-S7) for the character of Alcandro; it is an addition into Metastasio's original libretto. The aria "Se tu sprezzar pretendi" (A2-S2) included in the same manuscript is also written for a bass voice, yet it is highly problematical. It comes from his *La fida ninfa* (A2-S9) and seems like an insertion in the manuscript, with no connection to the rest of the opera. This makes us wonder whether therefore, in the whole corpus there is only one aria to be premiered in Italy by a bass, among the 1,392 (57.88% of the corpus) that were written for Italian theatres.

[94] It must be noted that, in opera, traditionally "bass roles were few and generally unimportant" and that, in the Metastasian repertoire, they had "only a modest role to play"; see Owen Jander, Lionel Sawkins, J. B. Steane, and Elizabeth Forbes, "Bass (Fr. basse; Ger. Bass; It. basso)", *Grove Music Online*, <https://www.oxfordmusiconline.com/> [consulted 23/03/2021]).

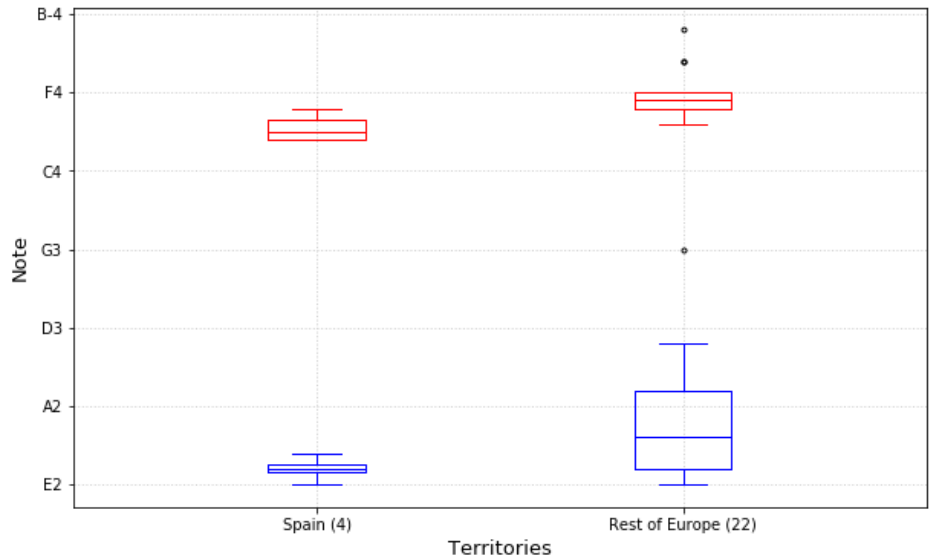

**Figure 10.** Compass of bass voices in Metastasian arias, 1730s-1750s: differences among territories. The boxplots represent the dispersion of the data.

Even though numbers may result tempting, they must however be placed into their context. It is true that the four arias for the Spanish court represent a 15.38% of the bass arias in our corpus and, given the relative weight of Spanish arias therein (3.70%), the Spanish sample for bass voices can be considered as representative. However, in the very few arias for bass (26) we observe a marked tendency towards higher voices from the 1760s onwards, perhaps due to the specific singers taking part in the premieres. Therefore, we have decided to focus the analysis on the 20 bass arias written in the 1730s, the 1740s, and the 1750s, as in these decades the treatment of the bass voices in our corpus is akin. For easier visualisation and interpretation of the data, boxplots such as Figure 10 are provided. They represent the dispersion of the absolute lowest —in blue— and the highest notes —in red— in the arias.[96] Within each of them, the central "boxes" (or rectangles) contain the 50% of the data, with the horizontal lines inside them representing the median, i.e., the middle note in the ordered set of either the absolute lowest or the absolute highest notes in the vocal parts in the relevant arias. The larger the height of the box, the more dispersed the notes are; in other words, the higher the box, the more varied the registral limits are for the select group of arias. The remaining notes extend between the box and the "whiskers", which determine, for each boxplot, the most extreme note that is *not* considered an outlying or an atypical value. Outliers are represented with crosses in the case of the lowest notes and with circles in the case of the highest ones.

As mentioned above, the bass singing in Corselli's 1744 *Achille in Sciro* was Antonio Montagnana, whose compass is supposed to have already shrunk at least to G2 to Eb4 by 1738.[97] However, five years later, in 1743, Corselli wrote for him the Christmas cantata *Ah del imperio*, where he reaches for three times a F#4. Therefore, the apparent deterioration of his voice does not seem to underlie Corselli's moderation regarding the treatment of the high extreme of his compass, particularly since Montagnana had to reach such a low note as E2, a note sung in other occasion only, by the bass premiering Aquilio's role in Caldara's *Adriano in Siria* (1732).[98]

---

[95] For more accurate computational processing of the data, international standards are used to refer to the octaves, with A4 being the central A (≈440Hz).

[96] The notes are represented in the y axis on a linear scale. The minuses between the note letters and the octave indices stand for flats (see, e.g., B-4 and A-5 in the axis labels), in consonance with the usual nomenclature in computational musicology.

[97] See Dean, "Montagnana, Antonio". In fact, in the previous decade he would sing up to a E2 ("Più bella, al tempo usato", A3-S2, in Caldara's *Adriano in Siria*, 1732).

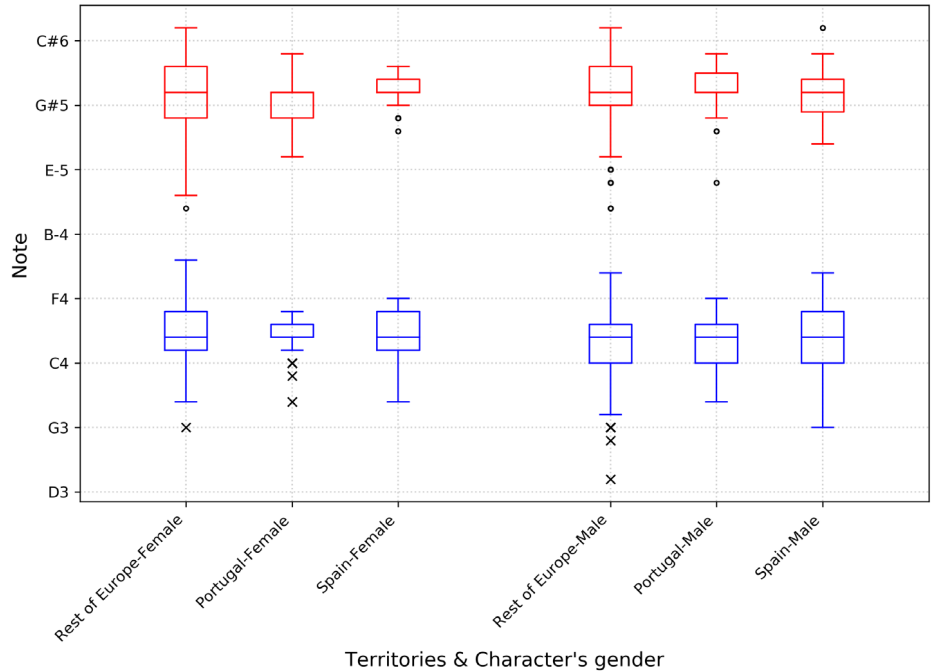

**Figure 11.** Compass of soprano voices depending on the characters' gender in Metastasian arias, 1740s-1760s: differences among territories. The boxplots represent the dispersion of the data.

The treatment of the soprano voices too must be approached cautiously. Throughout the corpus, there is a progressive increase in the compass towards the high register, this affecting both the lowest and the highest notes. As a consequence, we restrict the analysis to the soprano arias written in the 1740s, 1750s, 1760s, since, in this Metastasian corpus, median values for the extreme notes are comparable. In the 1,445 arias written in those three decades (60.58% of the corpus), it is evident that composers would exploit soprano compasses much more in what we have been calling "the rest of Europe", i.e., places other than Portugal and Spain, ranging from Eb3 up to D6.[99] In the Peninsula, they would never sing lower than G3 or, similarly, higher than D6.[100]

However, the ways in which such soprano voices were shaped depending on the characters' gender is thought-provoking. Whereas they tended to sing around the same notes (note the medians in Figure 11), the different lengths for the whiskers and the dispersion of the highest notes point to dissimilar vocal characterisations. Omitting outliers, it seems evident that, in the case of female characters, in the Peninsula and most markedly in Spain, their arias always went beyond a F#5. Additionally, only rarely did their arias surpassed the low limit of C#4.[101] The soprano arias for male characters show an opposite picture: whereas in the rest of Europe they could perfectly reach a D6, in the Iberian courts, and especially in Spain, they very rarely surpassed the B5.[102] Also, the low G3 are not statistically atypical. In other words, as a rule, female soprano arias in the Peninsula remain consistently high in register, whereas their male counterparts show a tendency towards the low part of the compass, similarly to how Motagnana's voice too was used.

---

[98] The score is edited in Torrente, *Fiesta de Navidad*, pp. 215-246.

[99] The Eb3 appears in the aria "Prudente mi chiedi?" (A2-S2) for Timante in Hasse's 1748 version of Metastasio's *Demofoonte*. The D6 features in 33 arias in this part of the corpus, including one that Conforto wrote for Naples ("Barbaro non comprendo" in the third act of his 1754 *Adriano nell'Indie*).

[100] The G3 corresponds to the aria "Se l'amor tuo mi rendi" (A3-S12) from Conforto's *Siroe*, and the D6 to "Ah, se in ciel benigne stelle" (A1-S2) from his *L'eroe cinese*.

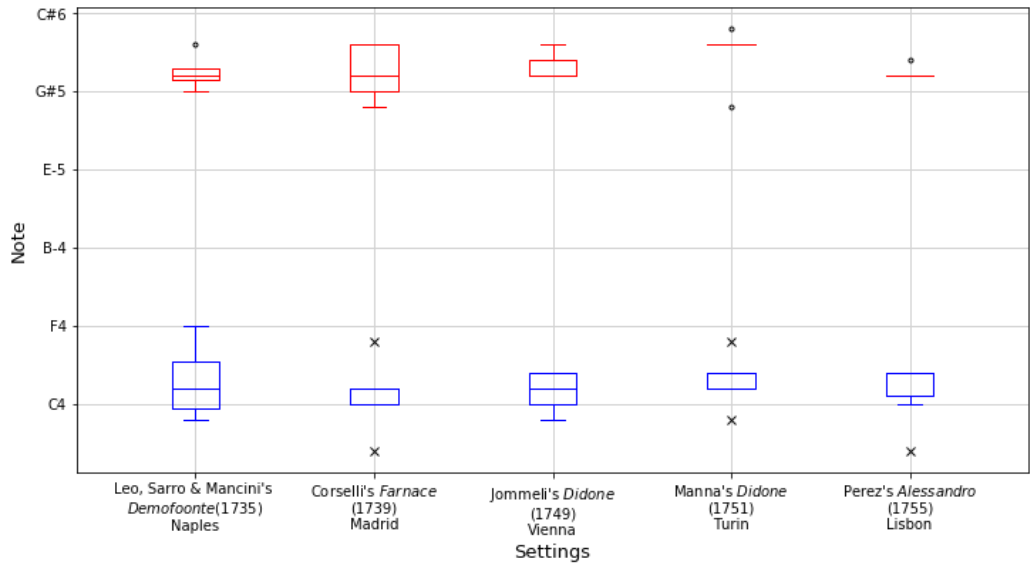

**Figure 12.** Caffarelli's compass, 1730s-1750s: differences among settings.
he boxplots represent the dispersion of the data.

Interestingly, in all of the Portuguese versions in our corpus both male and female characters were sung by male castrati and, therefore, we hypothesise that the divergent use of their voices was at least in part crafted with a view to musically differentiate among genders.[103] The case of the soprano Caffarelli, who in the versions included in our corpus sang male roles only, is highly illuminating in this respect. Even if, overall, his compass in the last setting, Perez's 1755 *Alessandro nell' Indie*,[104] is slightly more reduced than in previous versions elsewhere —his lowest notes there ranged from A3 to D4[105] and his highest notes always were a A5, except in one aria in which Perez made him reach a Bb5—[106], we cannot observe any significant shrinkage in his compass as the years elapsed. To illuminate this issue, we have included in the analysis a non-Metastasian court opera for Madrid, Corselli's *Farnace* (1739),[107] to bridge the gap between the years 1735 (Leo, Sarro, and Mancini's *Demofoonte*) and 1749 (Jommelli's *Didone abbandonata*). As Figure 12 shows, in the settings prior to Perez's *Alessandro*, and especially Manna's *Didone*, he had to reach higher notes. Also, the A3 that he sang in Perez's opera was not a unique instance, as he reached it in Corselli's *Farnace*. Importantly, the compass in the two parts that Händel wrote for him in London was B3 to A6, i.e., not as low as in the Peninsular scores.[108] In other words, again the composers writing for the Iberian courts seem to treat the high extreme of male characters' parts with caution while exploiting the low range.

---

[101] A3 in "Quando il mar bianccheggia e freme" (A2-S4), from Conforto's *L'eroe cinese*.

[102] "Ah, se in ciel benigne stelle" (A1-S2) and "Il mio dolor vedete" (A2-S2) from Conforto's *L'eroe cinese*.

[103] The moment in their careers naturally played a role too. As an example, the compass for Gioacchino Conti's arias in Italy in the 1740s (male principal roles from Galuppi's *Adriano* and Jommelli's *Artaserse*) is slightly higher (C4-C6) than in the Portuguese parts that Perez wrote for him (B3-B5) in 1752 and 1753 (in *Didone abbandonata*, *Demofoonte*, *L'eroe cinese*, and *Olimpiade*). Not surprisingly, the compass in the Italian settings correspond to his maximum amplitude, whereas the Portuguese roles are the last ones that he sang in his career. See Winton Dean, "Conti, Gioacchino ['Egizziello', 'Gizziello']", in *Grove Music Online*, <https://www.oxfordmusiconline.com/> [consulted: 23/03/2021].

[104] In fact, the last operas in which he sang were productions for the Portuguese court in 1755; see Winton Dean, "Caffarelli [Cafariello, Cafarellino, Gaffarello] [Majorano, Gaetano]", in *Grove Music Online*, <https://www.oxfordmusiconline.com/> [consulted: 23/03/2021].

[105] In "Destrier che all'armi usato" (A2-S10), on the one hand, and in the rest of the arias for the character of Poro in his *Alessandro nell'Indie*, on the other.

[106] In "Dov'è? S'affretti" (A3-S7) from his *Alessandro*.

[107] On a libretto by Antonio Maria Lucchini. We have used

|  | *p*-value |
|---|---|
| Aria form |  |
|     1740s | 0.045 |
|     1750s | **3.354·10⁻¹²** |
| Change of metre between A and B |  |
|     1740s | 0.566 |
|     1750s | 0.007 |
| Change of tempo between A and B |  |
|     1740s | 0.969 |
|     1750s | 0.471 |
| Change of key between A and B |  |
|     1740s | **2.496·10⁻⁹** |
|     1750s | **0.002** |

**Table 4.** *p*-values of the Chi-Squared Tests of Independence (CSTIs) for the contingency tables conformed by the three select territories (common to all tests: Spain, Portugal, and the rest of Europe) vs. each of the variables specified in the rows. Bold font indicates significance after Bonferroni correction (N = 26).

The same occurred in the case of Gioacchino Conti's. The compass for his arias in Italy in the 1740s (male principal roles from Galuppi's *Adriano* and Jommelli's *Artaserse*) is slightly higher (C4-C6) than in the Portuguese parts that Perez wrote for him (A3-B5) in 1752 and 1753 (in his *Didone abbandonata*, *Demofoonte*, *L'eroe cinese*, and *Olimpiade*).[109] Although the compass in the Italian settings correspond to his reported maximum amplitude, in the Portuguese roles, which furthermore were the last ones in his career, he had to sing an even lower note.[110]

---

the manuscript source E-Mmh Mus 679-682.

[108] See Dean: "Caffarelli".

[109] See Dean, "Conti, Gioacchino".

[110] We cannot wholly dismiss that these differences were a consequence of potentially different pitch standards; see Bruce Haynes, *A History of Performing Pitch: The Story of 'A'* (Lanham: Scarecrow Press, 2002).

\* \* \*

Springing from Italy, Metastasian *opera seria* was cultivated all throughout Europe, including the Iberian Peninsula. Whereas musicology has traditionally defended that the style was notably homogeneous, due in part to the composers and singers' circulation around the continent, this study has shown that some specific territorial, as well as personal, practices can nonetheless be distinguished.

David Perez apparently looked for balance in his settings of Metastasio's librettos for the Braganza court. On the one hand, he was respectful in the exploitation of the voices at his disposal, among which some of the most highly recognised names appear. Specially, he made the singers be at ease regarding the vocal compass in the arias, not yielding to any potential pressures for virtuoso display in the high region. In this regard, one can musically corroborate that these court productions were not subjected to potential virtuoso demands from the public. On the other hand, he was more experimental with the form, resorting almost only to the *dal segno* type and to some binary examples. He also experimented with changes of metre and tempo, and, most significantly, key and harmony. This latter point refers to the tonal contrast in the B section, as the move towards the subdominant strongly characterises his Portuguese settings. Similarly, the use of allegedly melancholic or non-bright keys, such as f minor and F major, seems to be a distinguishing trait of the Portuguese versions in the 1750s. For their part, composers for the Bourbons in Spain seemed to have looked for stability and courtly delight. They captured such traits in the form of binary metres, significantly C|, straightforward metric changes, stable moderate-to-fast tempi and, importantly, dominant-key relations, which could have reminded audiences of other repertoires well-known to them. See Table 4 and Tables 5a-c; note that the different tests were carried out on the basis of the investigated hypotheses as informed by a descriptive analysis of the data.

|  | Spain vs. rest of Europe & Portugal |
|---|---|
| Metre |  |
| Fewer simple duple time, 1740s | 0.079 |
| More C\|, 1740s | 0.067 |
| More C\|, 1750s | 0.028 |
| More 3/4, 1740s | 0.135 |
| Fewer 6/8, 1740s | 0.107 |
| Tempo |  |
| Mild fast tempi changes between A and B, 1750s | **0.001** |

**Table 5a.** *p*-values for the one-sided two-sample Test of Proportions (TPs) performed throughout the study: Spain vs. the rest of Europe, including Portugal (Portuguese data available for the 1750s only). Bold font indicates significance after Bonferroni correction (N = 26).

|  | Portugal vs. rest of Europe & Spain |
|---|---|
| Tempo |  |
| More changes of tempo and metre between A and B | **0** |
| Key |  |
| More F major | 0.186 |
| More Eb major | 0.006 |
| More E major | 0.003 |
| More f minor | **0** |
| Fewer C and A major among major-key arias | **0** |
| More B sections in the subdominant | **0.001** |
| Fewer B sections in the relative key | **0** |

**Table 5b.** *p*-values for the one-sided two-sample Test of Proportions (TPs) performed throughout the study: Portugal vs. the rest of Europe, including Spain, in the 1750s.

|  | Spain & Portugal vs. rest of Europe |
|---|---|
| Metre |  |
| More changes from simple duple to simple triple time between A and B | **0** |
| Tempo |  |
| More moderate tempi | 0.113 |
| More moderate or fast tempi | 0.010 |
| More changes of tempo between A and B | **0** |

**Table 5c.** *p*-values for the one-sided two-sample Test of Proportions (TPs) performed throughout the study: Spain and Portugal vs. the rest of Europe in the 1750s. Bold font indicates significance after Bonferroni correction (N = 26). Bold font indicates significance after Bonferroni correction (N = 26).

As a general Peninsular trait, it seems that the court audiences had, probably for different reasons, specific gender-related expectations. Composers purportedly tried to present on stage women that were not "masculine" —considering a low voice as "masculine"— and men that were, on the contrary, the most masculine ones in the corpus. This affected both sopranos, as Perez's treatment of Caffarelli and Gizziello's voices attests, and basses, who moreover were rare in the corpus. In the case of Portugal, this might have been dictated by a need of differentiation between male and female characters, who were always sung by male singers. In Spain, perhaps influenced by the stereotypes in the Spanish theatrical tradition, in which men and women played highly standardised roles,[111] Corselli, Conforto, Jommelli, and Galuppi as a group opted for highly characteristic approaches to the various voice types.

Of course, personal idiosyncrasies of composers and singers, such as the tendency towards particular musical features or the latters' vocal capacities at specific moments in their careers, played a role in shaping Iberian Metastasian *opera seria* in the eighteenth century. Yet the data available for this study indicate that compositional decisions nonetheless were partly shaped by local and court tastes. In this context, Leza's words that "Although they constituted a compact and recognisable model, the various elements of Italian opera nonetheless admit a fragmented reception that makes the study of its integration into other theatrical traditions especially relevant"[112] acquire fuller meaning beyond the literary adaptations and translations of Metastasio's texts.

Analysis of other Metastasian settings and of other genres cultivated in the Peninsula will be necessary to corroborate —or disprove— our conclusions. In any case, besides presenting some preliminary claims based on the statistical evaluation of a large corpus, this study has exemplified how pertinent Metastasian repertoire is as a departure point for constructing discourses on local compositional styles no matter their supposedly international character. This article has also shown that historical and aesthetic considerations must be taken into account when evaluating statistical results, such as the Spanish preference for *alla breve* settings, which on mathematical grounds may have been discarded as inconsequential. On the other hand, this article has illustrated how necessary hard data are in order to explore forgotten or less-known repertoires and how, without concrete data, hermeneutics can be no more than a lucubration lacking in solid grounds. "We could not do greater damage to opera of this era than by always considering it only *en bloc*, as if it was always identical, an art form of a single coinage. In the darkness of ignorance all cats seem grey".[113]

---

[111] This especially applies to spoken theatre; see Felipe B. Pedraza Jiménez and Milagros Rodríguez Cáceres, *Manual de literatura española. IV. Barroco: Teatro* (Tafalla: Cénlit, 1981), pp. 75-86. Further research is necessary to assess whether it is applicable to Spanish operas and *zarzuelas* besides the ones studied here. For an account of the characterising role of music and language in Spanish drama, see María Asunción Flórez, *Música teatral en el Madrid de los Austrias durante el Siglo de Oro* (Madrid: Instituto Complutense de Ciencias Musicales, 2006), pp. 124-134.

[112] Leza, "'Al dulce estilo de la culta Italia'", p. 308.
[113] Bianconi, "Le 'mutazioni sceniche' nel teatro d'opera", p. 74: "Non faremmo torto maggiore all'opera di quest'epoca che considerandola sempre solo *en bloc*, come fosse sempre identica, una forma d'arte tuta d'un sol conio. Nel buio dell'ignoranza finisce che tutti I gatti paion bigi".

**APPENDIX**

Settings of select *drammi per musica* by Metastasio and sources used in this study.

| Year | Opera | Composer | City | Source |
|---|---|---|---|---|
| 1724 | *Didone abbandonata* | D.N. Sarro | Naples | I-Nc 18.4.2 |
| 1726 | *Didone abbandonata* | L. Vinci | Rome | A-Wn Mus.Hs.17710 |
| 1730 | *Alessandro nell'Indie* | L. Vinci | Rome | I-Nc 32.4.9; D-Mbs Mus.ms. 169 |
| 1730 | *Artaserse* | L. Vinci | Rome | I-Nc 30.1.8; I-Vnm Mss.It.IV,244-246; A-Wn Mus.Hs.19120 |
| 1730 | *Didone abbandonata* | D.N. Sarro | Venice | I-Nc* 32.2.20 |
| 1731 | *Alessandro nell'Indie* | G.F. Häendel | London | GB-Lbl R.M.20.b.13 |
| 1731 | *Alessandro nell'Indie* | J.A. Hasse | Dresden | D-Mbs* Mus.ms. 213; D-Dl Mus.2477-F-9 |
| 1732 | *Adriano in Siria* | A. Caldara | Vienna | F-Pn L-3398 |
| 1733 | *Adriano in Siria* | G. Giacomelli | Venice | D-B Mus.ms. 7470 |
| 1733 | *Demofoonte* | A. Caldara | Vienna | A-Wn Mus.Hs.17107 |
| 1734 | *Adriano in Siria* | G.B. Pergolesi | Naples | I-Nc 30.4.10-11 |
| 1734 | *Demofoonte* | G.M. Schiassi | Venice | B-Bc 2355 |
| 1734 | *Olimpiade* | A. Vivaldi | Venice | I-Tf* Foà 39 |
| 1735 | *Adriano in Siria* | F.M. Veracini | London | D-Mbs Mus.ms. 143 |
| 1735 | *Demofonte* | L. Leo, D.N Sarro & F. Mancini | Naples | I-Nc 28.4.20 |
| 1735 | *Olimpiade* | G.B. Pergolesi | Rome | D-B Mus.ms. 17180; D-B Mus.ms. 142 |
| 1737 | *Adriano in Siria* | G.B. Ferrandini | Munich | D-Dl Mus.3037-F-1 |
| 1737 | *Olimpiade* | L. Leo | Naples | I-Nc 28.4.23 |
| 1738 | *Alessandro nell'Indie* | B. Galuppi | Mantua | F-Pn Mus. Ab. o. 156[1] |
| 1739 | *Adriano in Siria* | G.A. Ristori | Naples | I-Nc 31.6.23; D-Dl Mus.2455-F-8 |
| 1739 | *Artaserse* | G.B. Ferrandini | Munich | D-Dl Mus.3037-F-2; D-Dl Mus.3037-F-2 |
| 1739 | *Didone abbandonata* | G.B. Lampugnani | Padua | E-Mn M* 2369-2370 |
| 1740 | *Adriano in Siria* | M. Caballone | ? | F-Pn MS-2022 |
| 1740 | *Adriano in Siria* | B. Galuppi | Turin | B Bc 2093[2] |
| 1740 | *Artaserse* | J.A. Hasse | Dresden | D-LEu N.I.10308; D-Dl Mus.2477-F-2a/2b/2c; F-Pn D-5395 |
| 1741 | *Artaserse* | C.W. Gluck | Milan | F-Pn AB 0 147[3] |
| 1742 | *Didone abbandonata* | J.A. Hasse | Wermsdorf | I-Vnm Mss.It. IV,266; D-DI Mus. 2477-F-35&35a |

---

[1] Although, due to the first-page inscription, the manuscript is ascribed to Alessandri in the F-Pn catalogue, collation with other sources allowed us to determine that 21 out of the 28 arias in the source are by Galuppi, from both the 1738 and the 1755 versions.

[2] This source contains arias for both the 1740 and 1760 settings by Galuppi.

[3] The manuscript is catalogued as Zingarelli's version by F-Pn, yet the music is by Gluck, corresponding to his 1741 setting.

| Year | Opera | Composer | City | Source |
|------|-------|----------|------|--------|
| 1743 | *Demofoonte* | C.W. Gluck | Milan | B-Bc12801 |
| 1743 | *Demofoonte* | N. Jommelli | Padua | D-Sl* HB XVII 239 |
| 1744 | *Achille in Sciro* | F. Corselli | Madrid | E-Mbh Mus 677-679 |
| 1744 | *Alessandro nell'Indie* | C.H. Graun | Berlin | D-B* Am.B 191; F-Pn Mus. D-4997 |
| 1744 | *Didone abbandonata* | P. Scalabrini | Hamburg | I-MOe Mus. F. 1587 |
| 1746 | *Adriano in Siria* | C.H. Graun | Berlin | F-Pn D-5006 |
| 1746 | *Alessandro nell'Indie* | M. Caballone | Palermo | F-Pn MS-2021 |
| 1747 | *Didone abbandonata* | N. Jommelli | Rome | I-Nc 28.5.6 |
| 1748 | *Alessandro nell'Indie* | C.C. Wagenseil | Vienna | A-Wn Mus.Hs.18018; |
| 1748 | *Demofoonte* | J.A. Hasse | Dresden | D-Dl Mus.2477-F-53/53ª; D-LEm Becker III.15.12; D-HAmi MS 63 |
| 1749 | *Artaserse* | N. Jommelli | Rome | D Sl* HB XVII 234 |
| 1749 | *Artaserse* | D. Perez | Naples | I-Nc 30.4.1; I-Vnm Mss.It.IV,226-228 |
| 1749 | *Demofoonte* | B. Galuppi | Madrid | F-Pn* MS-1905; D-Dl Mus.2973-F-10,1&2&3; D-Dl Mus.2973-F-31,19 |
| 1749 | *Demofoonte* | J.A. Hasse | Venice | I-Vnm Mss.It.IV,247 |
| 1749 | *Didone abbandonata* | N. Jommelli | Vienna | A-Wn Mus.Hs.18282; D-Dl Mus.3032-F-1 |
| 1751 | *Adriano in Siria* | A. Adolfati | Genoa | F-Pn AB 0 161&162 |
| 1751 | *Artaserse* | N. Jommelli | Mannheim | D-B Mus.ms. 11245 |
| 1751 | *Demetrio* | N. Jommelli | Madrid | E-Mp Mus.mss.355-357 |
| 1751 | *Didone abbandonata* | G. Manna | Turin | D-MÜs SANT Hs 2469 |
| 1752 | *Adriano in Siria* | J.A. Hasse | Dresden | D-Leu N.I.10290a-c; D-Dl Mus.2477-F-67 |
| 1752 | *Demofoonte* | D. Perez | Lisbon | I-Vnm Mss.It.IV,232-234 |
| 1752 | *Didone abbandonata* | B. Galuppi | Madrid | B-Bc 2097 |
| 1752 | *Siroe* | N. Conforto | Madrid | D-Dl Mus.1-F-82,8-16; D-Dl Mus.3069-F-2,6 |
| 1753 | *Didone abbandonata* | N. Jommelli | Milan | I-Nc* 15.4.11; I-Nc Rari 7.7.15 |
| 1753 | *Didone abbandonata* | D. Perez | Lisbon | I-Vnm Mss.It.IV,214-216 |
| 1753 | *L'eroe ciñese* | D. Perez | Lisbon | I-Vnm Mss.It.IV,229-231 |
| 1753 | *Olimpiade* | D. Perez | Lisbon | I-Vnm Mss.It.IV,217-219 |
| 1754 | *Alessandro nell'Indie* | J.F. Agricola | Berlin | D-DS Mus.ms 10 |
| 1754 | *Adriano in Siria* | N. Conforto | Naples | I-Nc 25.2.32 |
| 1754 | *Adriano in Siria* | D. Perez | Lisbon | I-Vnm Mss.It.IV,220-222 |
| 1754 | *Demofoonte* | G. Manna | Turin | I-Nf* Inv.Nº.347; D-Dl Mus.1-F-82,16-4 |
| 1754 | *Ipermestra* | D. Perez | Lisbon | I-Vnm Mss.It.IV,579-581 |
| 1754 | *L'eroe cinese* | N. Conforto | Madrid | E-Mp MUS MSS 351-352 |
| 1755 | *Adriano in Siria* | A. Bernasconi | Munich | D-Mbs Mus.ms. 185 |
| 1755 | *Alessandro nell'Indie* | B. Galuppi | Venice | D-Mbs Mus.ms. 228 |
| 1755 | *Alessandro nell'Indie* | D. Perez | Lisbon | I-Vnm Mss.It.IV,223-225 |
| 1756 | *Artaserse* | N. Jommelli | Stuttgart | D-Sl HB XVII 730 |

| Year | Opera | Composer | City | Source |
|------|-------|----------|------|--------|
| 1756 | *Didone abbandonata* | A. Bernasconi | Munich | D-Mbs Mus.ms. 207; D-Mbs Mus.ms. 20890 |
| 1756 | *Nitteti* | N. Conforto | Madrid | P-La 44-V-49; I-Nc 26.6.1 |
| 1756 | *Olimpiade* | J.A. Hasse | Dresden | D-Dl Mus.2477-F-83; D-HAmi MS 83 |
| 1758 | *Alessandro nell'Indie* | N. Piccinni | Rome | I-Nc* 16.5.33-34 |
| 1758 | *Demofoonte* | A. Ferradini | Milan | I-Nc 27.6.15 |
| 1758 | *Demofoonte* | B. Galuppi | Padua | B-Bc* 2091; I-MOe Mus. F. 1313 |
| 1758 | *Demofoonte* | J.A. Hasse | Naples | D-Dl Mus.2477-F-57/57a |
| 1759 | *Adriano in Siria* | L. Borghi | Turin | I-Tf 10 V 19 |
| 1760 | *Adriano in Siria* | B. Galuppi | Venice | D-Dl Mus.2973-F-7 |
| 1760 | *Artaserse* | J.A. Hasse | Naples | I-Nc* 27.2.10-11; D-LEu N.I.10286; F-Pn D-2045-2047; F-Pn X-55-57 |
| 1761 | *Demofoonte* | N. Piccinni | Reggio Emilia | I-Nc* 15.1.9-10 |
| 1761 | *Olimpiade* | N. Jommelli | Stuttgart | F-Pn VM4-38 |
| 1762 | *Adriano in Siria* | D. Colla | Milan | I-Nc 25.2.27-29 |
| 1762 | *Adriano in Siria* | J.G. Schwanenberg | Brunswich | D-LEm Becker III.15.26 |
| 1762 | *Alessandro nell'Indie* | J.C. Bach | Naples | I-Nc 24.5.19; F-Pn Mus. D-360-361 |
| 1762 | *Alessandro nell'Indie* | T. Traetta | Reggio Emilia | D-B Mus.ms. 22004[4] |
| 1762 | *Didone abbandonata* | G. Sarti | Copenhagen | DK-Kk mu7502.0538 |
| 1763 | *Alessandro nell'Indie* | A. Sacchini | Venice | I-Nc 31.4.17-19[5] |
| 1763 | *Artaserse* | A. Bernasconi | Munich | D-Dl Mus.ms. 151; D-Mbs Mus.ms.190; F-Pn D-993-995 |
| 1763 | *Artaserse* | G. Scarlatti | Vienna | B-Bc 2354 |
| 1763 | *Didone abbandonata* | N. Jommelli | Stuttgart | A-Wn Mus.Hs.16488 |
| 1763 | *Didone abbandonata* | T. Traetta | Milan | I-Nc Rari Cornice 5.28-30 |
| 1764 | *Demofoonte* | N. Jommelli | Stuttgart | D-Sl HB XVII 240a-c |
| 1764 | *Didone abbandonata* | B. Galuppi | Venice | I-Nc 6.5.18-20 |
| 1764 | *Olimpiade* | A. Bernasconi | Munich | D-Mbs Mus.ms. 149; D-Mbs Mus.ms. 188 |
| 1764 | *Olimpiade* | F.L. Gassmann | Vienna | I-Nc 27.6.36-38 |
| 1766 | *Demetrio* | D. Perez | Lisbon | I-Nc 30.4.2-4 |
| 1766 | *Demofoonte* | A. Bernasconi | Munich | D-Mbs Mus.ms.184 |
| 1766 | *Alessandro nell'Indie* | G.F. di Majo | Mannheim | I-Nc 31.4.17-19[6] |

---

[4] E The manuscript exactly corresponds to Gatti's 1768 version, but has three arias ("Chi vive amante", "Voi che adorate", and "Mio ben ricordati") from Traetta's 1762 setting.

[5] Sacchini's authorship for some arias in this source has not been possible to ascertain. Such arias are treated as composed for a 1768 Neapolitan revival.

[6] The manuscript exactly corresponds to Sacchini's 1763 version, but has one addition ("Talor l'acceso") that is an aria from Majo's 1769 setting.

| Year | Opera | Composer | City | Source |
|---|---|---|---|---|
| b1766 | *Adriano in Siria* | G.B. Pescetti. | ? | B-Bc 12508; I-Bas IV 89-749 A [13-19] |
| 1768 | *Alessandro nell'Indie* | L. Gatti | Mantua | D-B Mus.ms. 22004 |
| 1768 | *Artaserse* | N. Piccinni | Naples | I-Nc 15.1.7-8; I-Nc 30.4.33-37 |
| 1768 | *Didone abbandonata* | A. Boroni | Prague | D-Dl Mus.3406-F-5 |
| 1768 | *Olimpiade* | N. Piccinni | Rome | I-Nc*15.1.12 |
| 1769 | *Adriano in Siria* | C.I. Monza | | I-Nc 29.5.12-14 |
| 1769 | *Alessandro nell'Indie* | L. Koželuch | Prague | A-Wn Mus.Hs.17792 |
| 1769 | *Didone abbandonata* | I. Celoniati | Milan | F-Pn D-1907&1908 |
| 1769 | *Didone abbandonata* | G.F. di Majo | Venice | D-Mbs Mus.ms. 20889 |
| 1769 | *Olimpiade* | P. Cafaro | Naples | F-Pn MS-1672 |
| 1770 | *Demofoonte* | N. Jommelli | Naples | I-Mc Noseda F.99.1-2; I-Nc 28.5.1-2; I-Nc 28.5.5; I-Nc 28.6.38-40 |
| 1770 | *Didone abbandonata* | N. Piccinni | Rome | I-Nc 16.4.27-28 |
| 1771 | *Artaserse* | G. Paisiello | Modena | I-Nc* 16.8.18-19 |
| 1771 | *Demofoonte* | G. Sarti | Copenhagen | DK-Kk mu7502.0838 |
| 1772 | *Alessandro nell'Indie* | P. Anfossi | Rome | F-Pn Mus. D. 97-99 |
| 1773 | *Alessandro nell'Indie* | G. Paisiello | Modena | I-Nc* Rari 3.5.6 |
| 1773 | *Demofoonte* | P. Anfossi | Rome | D-MÜs SANT Hs 140a-c |
| 1773 | *Didone abbandonata* | D. Colla | Turin | I-Tf 1 VII 4-6 |
| 1774 | *Alessandro nell'Indie* | N. Piccinni | Naples | I-Nc 30.4.30-32 |
| 1774 | *Artaserse* | J. Mysliveček | Naples | I-Nc 29.4.32-34 |
| 1774 | *Olimpiade* | N. Piccinni | Naples | I-Nc* 15.1.11; I-Nc 30.3.25-27 |
| 1775 | *Olimpiade* | P. Anfossi | Venice | F-Fn D-202&203 |
| 1775 | *Demofoonte* | J. Mysliveček | Naples | A-Wn* Mus.Hs.16421; I-Nc 29.3.7-9 |
| 1776 | *Artaserse* | F. Bertoni | Forli | I-Mc Part. Tr. ms. 25 |
| 1776 | *Didone abbandonata* | J. Schuster | Naples | D-Dl* Mus.3549-F-10 |
| 1778 | *Adriano in Siria* | F. Alessandri | Venice | D-Mbs Mus.ms. 521 |
| 1778 | *Alessandro nell'Indie* | M. Mortellari | Siena | I-Mc Part. Tr. ms. 252 |
| 1778 | *Olimpiade* | J. Mysliveček | Naples | I-Nc 29.3.15-17 |
| 1780 | *Artaserse* | L. Caruso | Florence | I-Bc EE.28 |
| 1780 | *Didone abbansonata* | G. Astarita | Bratislava | A-Wn Mus.Hs.16538 |
| 1781 | *Alessandro nell'Indie* | D. Cimarosa | Rome | I-Nc* 13.3.11-12 |
| 1781 | *Olimpiade* | F. Bianchi | Milan | F-Pn D-1081&1082 |
| 1782 | *Adriano in Siria* | W. Rust | Turin | I-Tf 1 VI 16-18 |
| 1783 | *Artaserse* | F. Alessandri | Naples | I-Nc* 24.6.3-4 |
| 1784 | *Artaserse* | D. Cimarosa | Turin | I-Nc* 13.3.16-17 |

| Year | Opera | Composer | City | Source |
|------|-------|----------|------|--------|
| 1784 | *Olimpiade* | D. Cimarosa | Vicenza | I-Nc+ 14.8.16-17 |
| 1785 | *Alessandro nell'Indie* | F. Bianchi | Venice | I-Bc DD.165 (1-3) |
| 1786 | *Olimpiade* | G. Paisiello | Naples | I-Fc B.I.78; D-Mbs Mus.ms. 539 |
| 1788 | *Artaserse* | A. Tarchi | Mantua | D-Mbs Mus.ms. 547 |
| 1789 | *Alessandro nell'Indie* | P.A. Guglielmi | Naples | I-Nc 27.4.26 |
| 1794 | *Didone abbandonata* | G. Paisiello | Naples | I-Nc 16.8.36-37 |
| 1810 | *Didone abbandonata* | F. Paër | Paris | I-MOe Mus. F. 861 |